\documentclass[12pt,preprint]{aastex}

\newcommand{\et}{et al.}
\newcommand{\kms}{km s$^{-1}$}
\newcommand{\ha}{H$\alpha$}
\newcommand{\solar}{\ifmmode_{\sun}\else$_{\sun}$\fi}
\newcommand{\msun}{M$_{\sun}$}
\newcommand{\ergsec}{ergs s$^{-1}$}
\newcommand{\HII}{H$\,${\sc ii}}
\newcommand{\HI}{H$\,${\sc i}}
\newcommand{\coldens}{atoms cm$^{-2}$}
\newcommand{\microns}{$\mu$m}
\newcommand{\nstar}{N$_{\star}$}
\newcommand{\cd}{C$+$D}

\begin{document}

\slugcomment{To appear in AJ}

\title{DDO 88: A Galaxy-Sized Hole in the Interstellar Medium}

\author{Caroline E.\ Simpson}
\affil{Department of Physics, Florida International University, 
University Park, Miami, Florida 33199 USA}
\email{simpsonc@galaxy.fiu.edu}

\author{Deidre A.\ Hunter}
\affil{Lowell Observatory, 1400 West Mars Hill Road, Flagstaff, Arizona 86001
USA}
\email{dah@lowell.edu}

\and

\author{Patricia M.\ Knezek} \affil{WIYN Consortium, Inc., 950 N.\
Cherry Ave., Tucson, Arizona 85726-6732 USA} \email{knezek@noao.edu}

\begin{abstract}

We present an \HI\ and optical study of the gas-rich dwarf irregular
galaxy DDO 88. Although DDO 88's global optical and \HI\ parameters
are normal for its morphological type, it hosts a large (3 kpc
diameter) and unusually complete ring of enhanced \HI\ emission.  The
normal appearance of this galaxy in the optical and the outer regions
of the \HI\ give no hint of the presence of the striking \HI\ ring in
the inner regions.  The gas ring is located at approximately one-third
of the total \HI\ radius and one-half the optically-defined Holmberg
radius, and contains 30\% of the total \HI\ of the galaxy.  The ring
surrounds a central depression in the \HI\ distribution.  If the \HI\
ring and central depression in the gas were formed by the energy input
from winds and supernova 
explosions of massive stars formed in a starburst, the star-forming
event would have formed 0.1--1\% of the total stellar mass of the
galaxy. However, the UBV colors in the \HI\ hole are not bluer than
the rest of the galaxy as would be expected if an unusual star-forming
event had taken place there recently, but there is an old ($\sim$ 1--3 
Gyr), red cluster near the center of the hole that is massive enough to
have produced the hole in the \HI. An age estimate for the ring,
however, is uncertain because it is not observed to be expanding.  An
expansion model produces a lower estimate of 0.5 Gyr, but the presence
of faint star formation regions associated with the ring indicate a much
younger age. We also estimate that the ring could have dispersed by now
if it is older than 0.5 Gyr. This implies that the ring
is younger than 0.5 Gyr. A younger age would indicate that the red
cluster did not produce the hole and ring. Therefore, uncertainties
prevent us from concluding that the cluster and the \HI\ hole are
definitely related. If this ring and the depression in the gas which it
surrounds were not formed by stellar winds and supernovae, this would
indicate that some other, currently unidentified, mechanism is operating.

\end{abstract}


\section{Introduction}

DDO 88 (NGC 3377A; UGC 5889), is a gas-rich dwarf irregular galaxy
located in the Leo I group (M96).  It came to our attention when 21 cm
observations taken with the Very Large Array (VLA) Radio
Telescope\footnote{The VLA is a facility of the  
National Radio Astronomy Observatory, operated by Associated
Universities, Inc., under cooperative agreement with the National
Science Foundation.} indicated the presence of an apparent ring of
\HI\ surrounding the optical body of the galaxy (Simpson \& Gottesman
2000). Although these data suffered from a lack of sensitivity, the
suggestion of a large hole ($\sim$ 2--3 kpc; D = 7.4 Mpc) in the \HI\
distribution was intriguing. There are examples of other large holes
in similar galaxies (Walter \& Brinks 1999; Wilcots \& Miller 1998;
Puche \et\ 1992; Sargent \et\ 1983), but this seemed extreme.

Star formation episodes are thought to more strongly affect the
interstellar medium (ISM) in low mass galaxies and have been invoked
to explain the presence of shells, holes, and filaments observed in
small galaxies (e.g.\ Ott \et\ 2001; Stewart \et\ 2000; Walter \& Brinks
1999; Martin 
1998; Meurer \et\ 1998; Young \& Lo 1996; Meurer \et\ 1992; Puche \et\
1992; but see Rhode \et\ 1999).  Star formation in low mass, slowly
rotating galaxies appears to be regulated by the feedback between the
energy deposited in the ISM by massive star formation and the ISM
itself. Energy transferred from stellar winds and supernovae events
both heats up and disperses the ISM surrounding the
star formation area (e.g.\ Scalo \& Chappell 1999). It is thought that
as the \HI\ column density drops below a critical threshold neccessary
for star formation to occur, the star formation episode will switch
off. Meanwhile, cold, dense material is swept up and compressed around
the edges of the region, triggering further star formation. This
effect should be particularly strong on low-mass systems such as
dwarfs (Mac Low \& Ferrara 1999; de Young \& Heckman 1994).

To further investigate the interplay between star formation and the
\HI\ reservoir in irregular galaxies, we have obtained additional
radio and optical observations of DDO 88. High sensitivity VLA D-array
and high resolution CS-array 21 cm observations have been taken, along
with U, B, V, R, H, and \ha\ images, as well as long slit spectra. We
present here the results from this new \HI\ and optical study. 

\section{Observations} 

\subsection{HI} 
DDO 88 was observed using the D-array configuration of the VLA for a
total of approximately 7.5 hours over two days and for approximately 2.5
hours using the CS-array configuration. Table~\ref{tabhi} lists all the
observational parameters for each data set. The D-array provides numerous
short spacings resulting in high sensitivity but low (nominally
45\arcsec) resolution, while the CS-array provides longer baselines
and therefore higher resolution (nominally 12.5\arcsec), but with some
loss in sensitivity.\footnote{The CS-array was a modified version of
the C-array, with one antenna moved from the middle of the north arm
to a central pad to provide some short spacings. It has since replaced
the original C-array.} All observations
were made using a 128 channel spectrometer with a 
channel separation of 12.2 kHz (2.6 \kms) resulting in a total
bandwidth of 1.56 MHz. The 2IF mode was used with on-line Hanning
smoothing. The nearby continuum source 1117+146 was used as a phase
calibrator, with 1328+307 (B1950) used as a flux and bandpass
calibrator for the D-array observations, and 0538+498 (J2000) used for the
CS-array observations.

\placetable{tabhi}

The observations were edited and calibrated using
the standard routines in {\sc aips}. There was solar interference present in
all the data sets, primarily affecting the short baselines. The D
array observations were the most affected, with up to 20\% of the
data contaminated. For both the D and CS data the contaminated
visibilities were flagged, and satisfactory solutions were
subsequently obtained. The continuum emission was removed in the {\it
u-v} plane using an average of emission-free channels on both sides
of the galaxy spectrum.

The D-array data sets from each day were edited and calibrated separately, then
combined in the {\it u-v} plane and ourier-transformed using the {\sc
aips} task {\sc imagr} to make a cube of images; one image for each
frequency channel observed. Natural weighting was used to produce
images with the greatest sensitivity to low-level emission. This image
cube has a spatial resolution of $55.59\arcsec\times 53.56\arcsec$, with a
single channel r.m.s.\ ($\sigma$) of 0.7 mJy/B. During imaging, each
channel was {\sc clean}ed (Clark 1980) to this level to remove the
effects of the dirty beam. Channels containing signal in the resulting
data cube were then integrated in velocity using the AIPS task {\sc
momnt}, which smooths the data spatially and in velocity before
applying a conditional blanking to each channel. Pixels in the
original (unsmoothed) cube that correspond to pixels in the smoothed
cube with values below a designated level (in this case $1.75\sigma$)
are blanked. The blanked cube, now containing only significant
emission, is then integrated to produce three moment maps: the first
is the integrated \HI\ flux, the second is the temperature-weighted
velocity field, and the 
third is a map of the velocity dispersions associated with each pixel.

When imaging the CS-array data, various weighting schemes were tested
to find the optimal combination of resolution (beam size) and
sensitivity (single-channel r.m.s.). Uniform weighting with a {\sc
robust} factor of $-1$ produced images with the least degradation in
sensitivity (single-channel noise increased by 1.5) for the best
increase in resolution (beam size decreased by 1.6). This high
resolution cube has a spatial resolution of $14.5\arcsec \times
13.1$\arcsec\ and a single-channel r.m.s.\ of 1.26 mJy/B. As with the
D-array data, the CS-array data were {\sc clean}ed to this $1 \sigma$
level, and {\sc momnt} was used to apply a conditional blanking (using
a value of $1.4 \sigma$ as a cutoff) and to integrate the cube to
produce moment maps.

Finally, the edited, calibrated, and continuum-subtracted D and
CS-array data sets were combined in the {\it u-v} plane to produce a
data set with both good resolution ($16.8\arcsec\times14.6$\arcsec)
and sensitivity (0.98 mJy/B). This ``C$+$D''-array data set was then
fourier-transformed and {\sc clean}ed to the $1\sigma$ level using
{\sc imagr}. Uniform weighting with a {\sc robust} factor of $-1$ was
used. Tests were run to determine the optimum value for the
integration flux cutoff in {\sc momnt}, and moment maps were produced
using a cut-off value of $1.25\sigma$.

\begin{figure}
\plotone{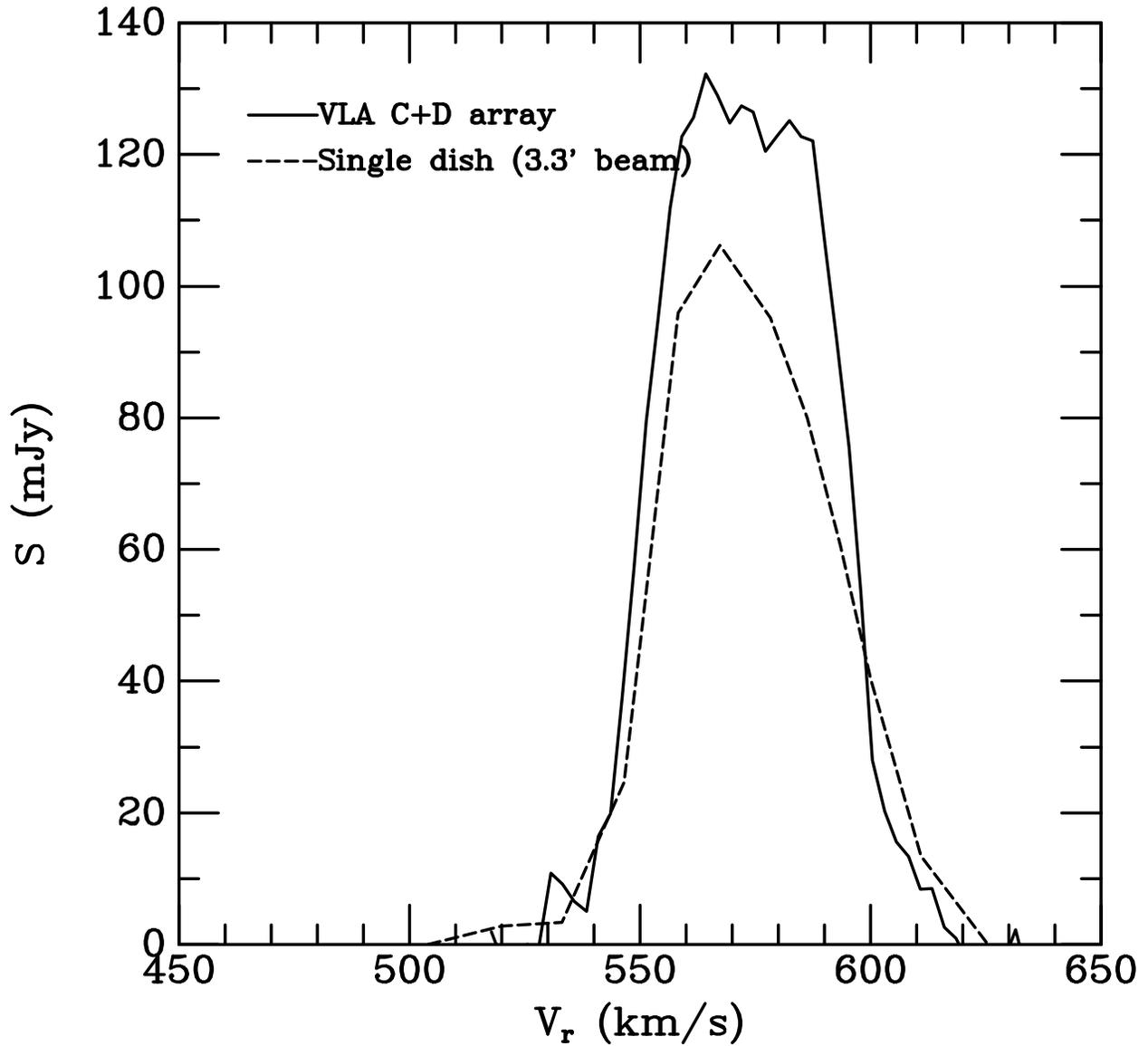}
\caption{Integrated velocity profile made by summing flux over a square
$\sim$3\protect\arcmin\ on a side from the \protect\cd\ array data.
This is compared to the profile obtained with a singledish radio
telescope with a 3.3\protect\arcmin\ beam by Schneider \et\ (1990).
\label{figsingle}}
\end{figure}

The flux of the \cd-array data cube was integrated in each channel
containing signal over a square 184\arcsec\ on a side, encompassing
the area over which \HI\ is detected in the integrated moment map.  The
velocity profile is shown and compared to the single-dish observation
of Schneider \et\ (1990) in Figure~\ref{figsingle}.  The single-dish
observation was made with the Arecibo telescope and a 3.3\arcmin\
beam. Thus, the Arecibo beam is comparable to the size of the \HI\
distribution that we detect.  Our VLA flux profile has a peak that is
15\% higher than the peak from the single-dish data.  However, the
total integrated flux is actually 5\% lower than the corrected flux
quoted by Schneider \et\ The integrated flux from our D-array data
alone, which is more sensitive to lower column density emission, is
6.04 Jy \kms, which is 16\% higher than the flux given by Schneider
\et. Therefore, it seems 
that we have recovered all of the flux detected by Arecibo, 
and perhaps a bit more.

\placefigure{figsingle}

\subsection{Optical images} 

We obtained UBV images of DDO 88 using the Lowell Observatory 1.1 m
Hall telescope March 1999 and April 2000.  The detector was a SITe
2048$\times$2048 CCD.  The telescope position was offset 20\arcsec\
between the three images taken in each filter in order to improve the
final flat-fielding.  Exposure times of 1800 s, 1800 s, and 1200 s for
UBV, respectively, were used.  The electronic pedestal was subtracted
using the overscan strip, and the images were flat-fielded using sky
flats.  Landolt (1992) standard stars were used to calibrate the
photometry.  The pixel scale was 1.13\arcsec; the seeing was
$\sim$3\arcsec.  For doing surface photometry on the DDO 88 images,
foreground stars and background galaxies were edited out of the
images and the background sky was fit with a two-dimensional Legendre
function and subtracted.

We also obtained \ha\ images of DDO 88 during April 1995 with the
Perkins 1.8 m telescope at Lowell Observatory. The observations used
an 800$\times$800 TI CCD provided to Lowell Observatory by the
National Science Foundation, the Ohio State University Fabry-Perot
that was used simply as a 3:1 focal reducer, an \ha\ filter with a
FWHM of 32 \AA, and an off-band filter centered at 6440 \AA\ with a
FWHM of 95 \AA .  The off-band filter was used to subtract stellar
continuum from the \ha\ filter to leave only \ha\ nebular emission.
The \ha\ exposure was 1800 s.  The pixel scale was 0.49\arcsec, and
the seeing was $\sim$2.3\arcsec.  The \ha\ emission was calibrated
using five spectrophotometric standard stars with minimal \ha\
absorption features (Stone 1977; Oke \& Gunn 1983).  The \ha\ filter
was also calibrated on other observing runs with spectrophotometry of
the \HII\ regions NGC 2363 (Kennicutt, Balick, \& Heckman 1980) and NGC
604 (Hunter \& Elmegreen 2004), and the \HII\ region calibration agreed
with the spectrophotometric standard star calibration to 4\%.  The
\ha\ photometry was corrected for the change in band-pass with
temperature and [NII] contamination, although these corrections were
small. A standard Anderson Mesa atmospheric extinction coefficient was
assumed for \ha.

The H band data were taken on February 28, 2003 at the WIYN Observatory
3.5 m telescope at Kitt Peak National Observatory.  The Near-Infrared 
Imager (NIRIM) was used (Mexiner, Young Owl, \& Leach 1999) 
with a plate scale of 0.69\arcsec/pixel, and a field of view of
$\sim$2.9\arcsec$\times$2.9\arcsec.  
The data were dark-subtracted and flat-fielded (using dome 
flats). Source and sky images were taken in an off-on-on-off repeating pattern
to enable the removal of atmospheric background emission.  The median of the 
three or four sky images closest in time was subtracted from each on-source 
image. The individual on-source frames were shifted to align to the 
nearest half pixel, and combined.  This resulted in a total of 15 minutes 
on-source integration.  The data were calibrated using the only two stars 
on the image bright enough to be present in the
2MASS catalog\footnote{\url{http://www.ipac.caltech.edu/2mass}}, and thus we  
estimate that the magnitudes are accurate to only $\sim$30\%.

In addition to the UBVH and \ha\ images presented here, B, R, and \ha\
images of DDO 88 obtained with the Michigan-Dartmouth-MIT Observatory
1.3 m telescope were discussed by Knezek, Sembach, \& Gallagher
(1999). We employ those data here as well. See Knezek \et\ for details
on the reduction of those data.

\subsection{Optical spectroscopy} 

Long-slit spectra were obtained of DDO 88 along three position angles 
with the Double-Imaging Spectrograph on the 3.5 m telescope at Apache
Point Observatory 20-22 1998 March. The spectrograph uses a
dichroic and two independent collimators and cameras to allow one
to obtain a blue and a red spectrum simultaneously.
There were 4 grating setups used to observe DDO 88, and they
have the following wavelength coverages and dispersions:
1) blue: 4360--5090 \AA, 1.55 \AA/pixel;
   red:  5890--6915 \AA, 1.3 \AA/pixel;
2) blue: 3800--4680 \AA, 1.55 \AA/pixel;
   red:  4500--6230 \AA, 1.3 \AA/pixel;
3) blue: 3600--6420 \AA, 6.3 \AA/pixel;
   red:  3600--8470 \AA, 7.2 \AA/pixel;
and
4) blue: 3600--6420 \AA, 6.3 \AA/pixel;
   red:  4300--9800 \AA, 7.2 \AA/pixel.
A 1.5\arcsec\ slit was used at position angles of 155.5\arcdeg,
51.5\arcdeg, and 124\arcdeg. Spectrophotometric standard stars
were observed to determine the response function, and HeNeAr arc
lamps were used to set the wavelength scale.

The blue camera used a 512$\times$512 SITe CCD with a scale of
1.1\arcsec/pixel and the red camera used
an 800$\times$800 TI CCD with a scale of 0.6\arcsec/pixel. 
The electronic pedestal was subtracted
from the two-dimensional images, a bias frame was subtracted, and 
pixel-to-pixel sensitivity variations were removed using observations
of internal continuum lamps.
The arc lamps were traced along the slit to map
spatial distortions and the images were linearly repixelized in wavelength.
The two-dimensional images were corrected for extinction using
the standard extinction function from Kitt Peak National Observatory,
they were fluxed using the sensitivity function determined from
the standard stars, multiple images were combined
to remove cosmic rays, and background sky was fit and subtracted using
the part of the slit beyond emission from DDO 88.
Finally, the \HII\ regions along the slit were identified, and one-dimensional
spectra were extracted. 

The emission lines in each one-dimensional spectrum were fit with Gaussians
to determine the flux. The signal-to-noise of a line was taken as
the rms of the continuum surrounding the emission line times the FWHM 
of the line. Several spectra were smoothed to increase the signal to
noise. Since there is more than one grating setting for each slit
position, there are often multiple measurements of a line. In addition,
two \HII\ regions were captured on two slit positions. Therefore, the
emission-line flux ratios were taken as the average. 

Emission-line ratios are given in Table~\ref{tabhii}. The ratio of 
[OIII]$\lambda$5007/[OIII]$\lambda$4959 is one diagnostic of the quality
of the data. For a nebula with an electron density of 100 cm$^{-2}$ and
a nebular temperature of 10$^4$ K, this ratio should be 2.9. For our
\HII\ regions, the ratio is within 1$\sigma$ of the expected value.  For
these same nebular parameters, we determine the E(B$-$V) from the ratio
of \ha\ to H$\beta$, using the Cardelli, Clayton, \& Mathis
(1989) reddening function.  These values range from 0.03$\pm$0.08 to
0.16$\pm$0.10, where foreground E(B$-$V)$_f$ is 0.01.
[OII]$\lambda$3727 was measured only for one \HII\ region.  The method
by McGaugh (1991) for determining the oxygen abundance from
[OIII]$\lambda$5007,4959 and [OII]$\lambda$3727 yields a value of 7.9
for 12$+$log(O/H) if we assume the lower abundance branch and 8.8 if we
assume the higher abundance branch. A value of 7.9 would place DDO 88 at
the low abundance end of the range for Im galaxies, and a value of 8.8
would be higher than most Im galaxies (Hunter \& Hoffman 1999).

\placetable{tabhii}

\section{Results: Optical} 

\subsection{General morphology}

\begin{figure}
\plotone{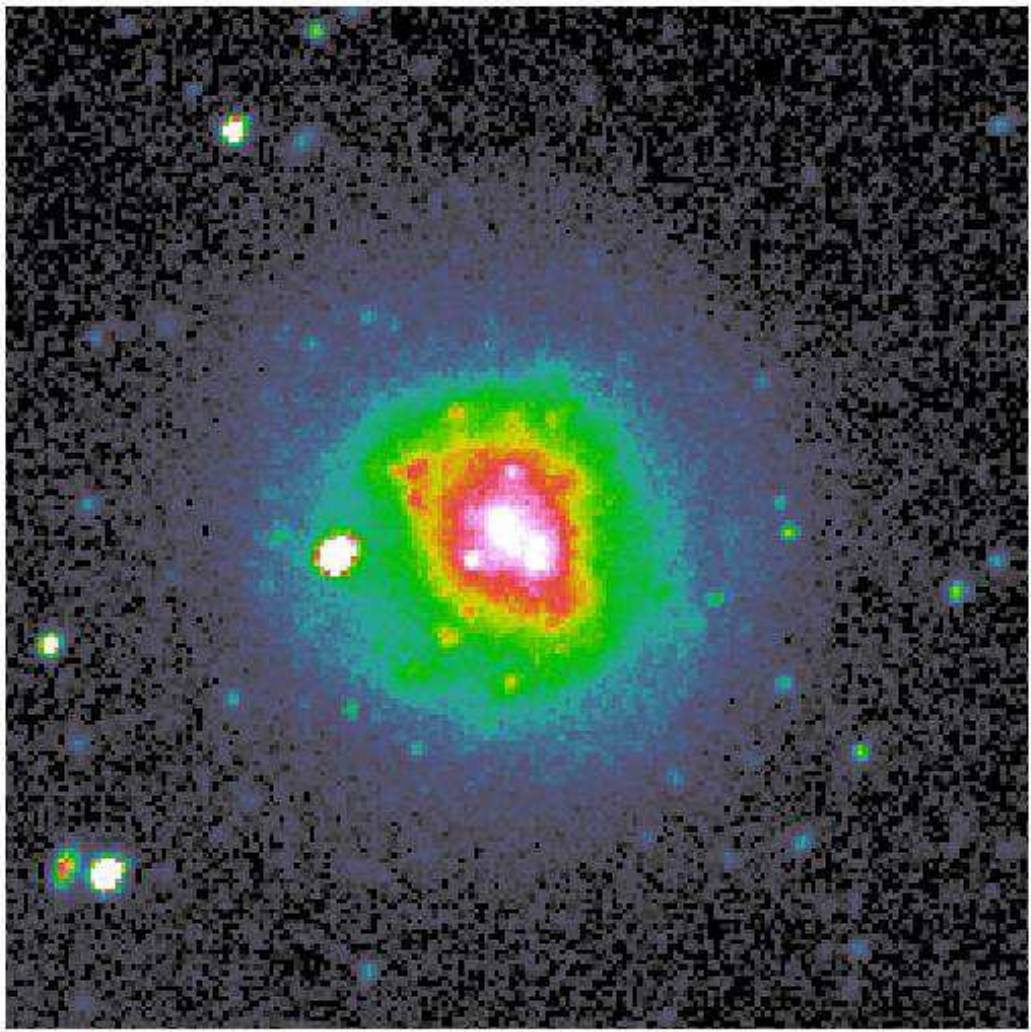}
\caption{False-color representation of the V-band image of DDO 88. 
We show the logarithm of the V-band image in order to allow
comparison of the inner and outer parts of the galaxy. North is up; east
is to the left.
\label{figv}}
\end{figure}

Our V-band image of DDO 88 is shown in Figure \ref{figv}.  We display
the logarithm of the image in a false-color representation so that the
reader can see the bright inner parts and the fainter outer parts in
one image.  We see that DDO 88 changes its shape slightly as it goes
from the inner to the outer galaxy, becoming more round in the outer
parts.  Specifically, we find an ellipticity ($1 - b/a$) from the
minor-to-major axis ratio $b/a$ of 0.13 in the inner galaxy at a
V-band surface brightness corrected for reddening
$\mu_{V,0}$ of 23.7 mag to 24.7 mag and an
ellipticity of 0.03 in the outer parts at a $\mu_{V,0}$ of 26.2 mag.
If the intrinsic $(b/a)_0$ is 0.3 (Hodge \& Hitchcock 1966), 
as is often taken for irregulars, the
observed outer $b/a$ implies an inclination of 16\arcdeg. Thus, the
galaxy is observed close to face-on.

\placefigure{figv}

In the V-band image we can also see an asymmetry at low surface
brightness levels. The galaxy appears to be slightly more extended to
the north and northwest than to the south and southeast.  However,
this kind of optical asymmetry is seen in other irregular galaxies
(see, for example, NGC 2366; Hunter, Elmegreen, \& van Woerden 2001).
DDO 88 is located only 15 kpc on the plane of the sky and 93 \kms\
radially from the giant elliptical galaxy NGC 3377 (hence, DDO 88's
alternate name of NGC 3377A).  However, except for this small optical
asymmetry, we see no signs of disturbance in DDO 88 that could be
ascribed to an interaction with this much larger galaxy (NGC
3377 is about 3 mag brighter in B than DDO 88).

\begin{figure}
\plotone{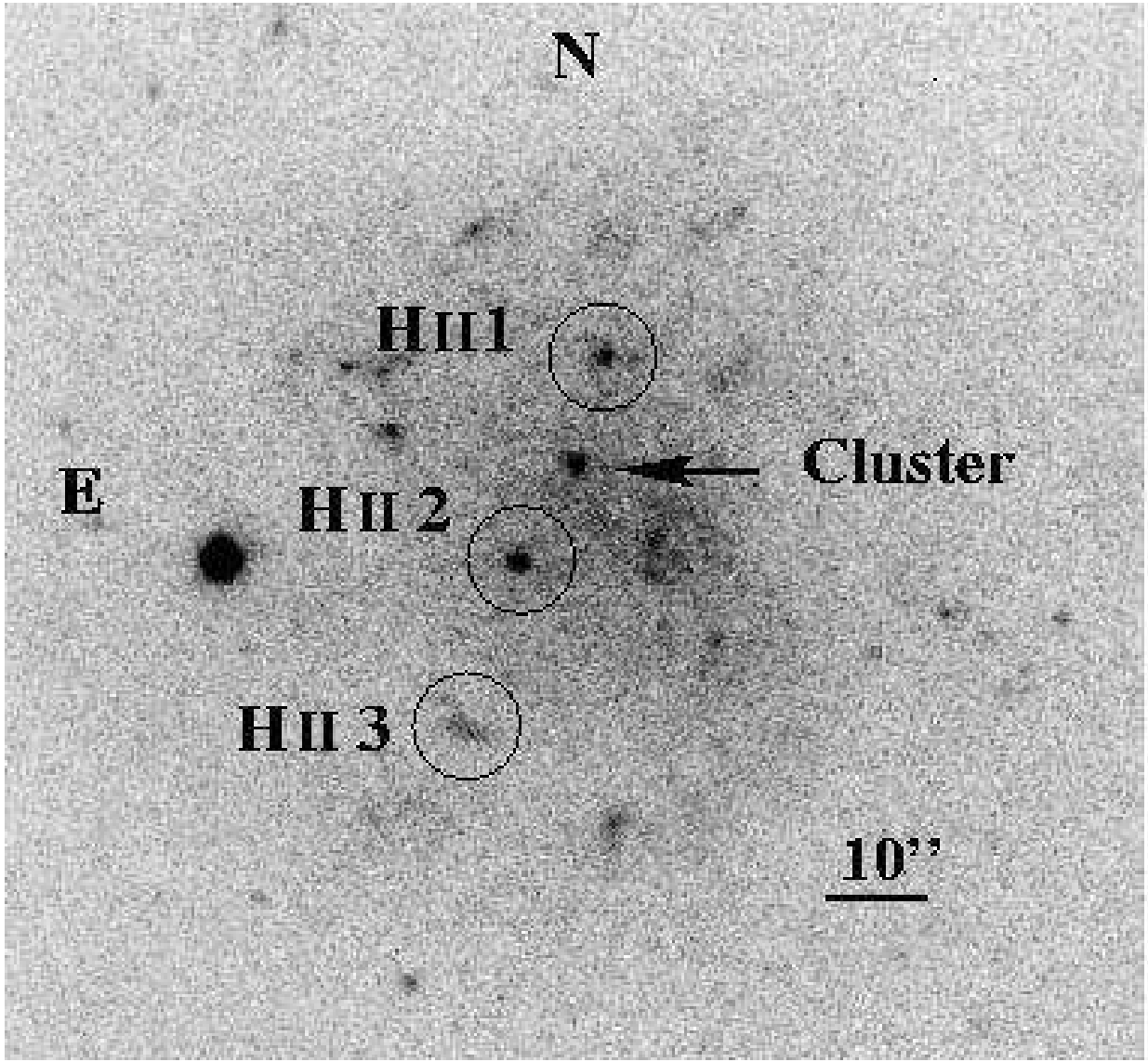}
\caption{Photographic plate image from the Carnegie Atlas (Sandage \&
Bedke 1994) taken with the Palomar 200-inch using a 103aO+GG1 emulsion
(4050\protect\AA).  The cluster is located in the center of the galaxy, in
between two \protect\HII\ regions. North is up, east is to the left.
\label{figbedke}}
\end{figure}

There is a red object located in the center of the galaxy that is
visible in the H-band image. Radial profiles on both the V and H images
show that it is extended, so it is probably a star cluster. It is the
center of the three bright regions extending from the northwest to the
southeast in Figure~\ref{figbedke}\footnote{Digitized image obtained
from NED. \url{http://nedwww.ipac.caltech.edu}.}. The other two bright
regions bracketing the cluster are \HII\ regions 1 and 2 in
Table~\ref{tabhii}. Region 2, just to the southeast of the cluster, also
shows up in the H image; it is the only \HII\ region detected in H. We
have performed multi-aperture photometry of both the cluster and the
southeast \HII\ region which is discussed below.

\placefigure{figbedke}

\subsection{Surface Photometry}

\begin{figure}
\epsscale{0.75}
\plotone{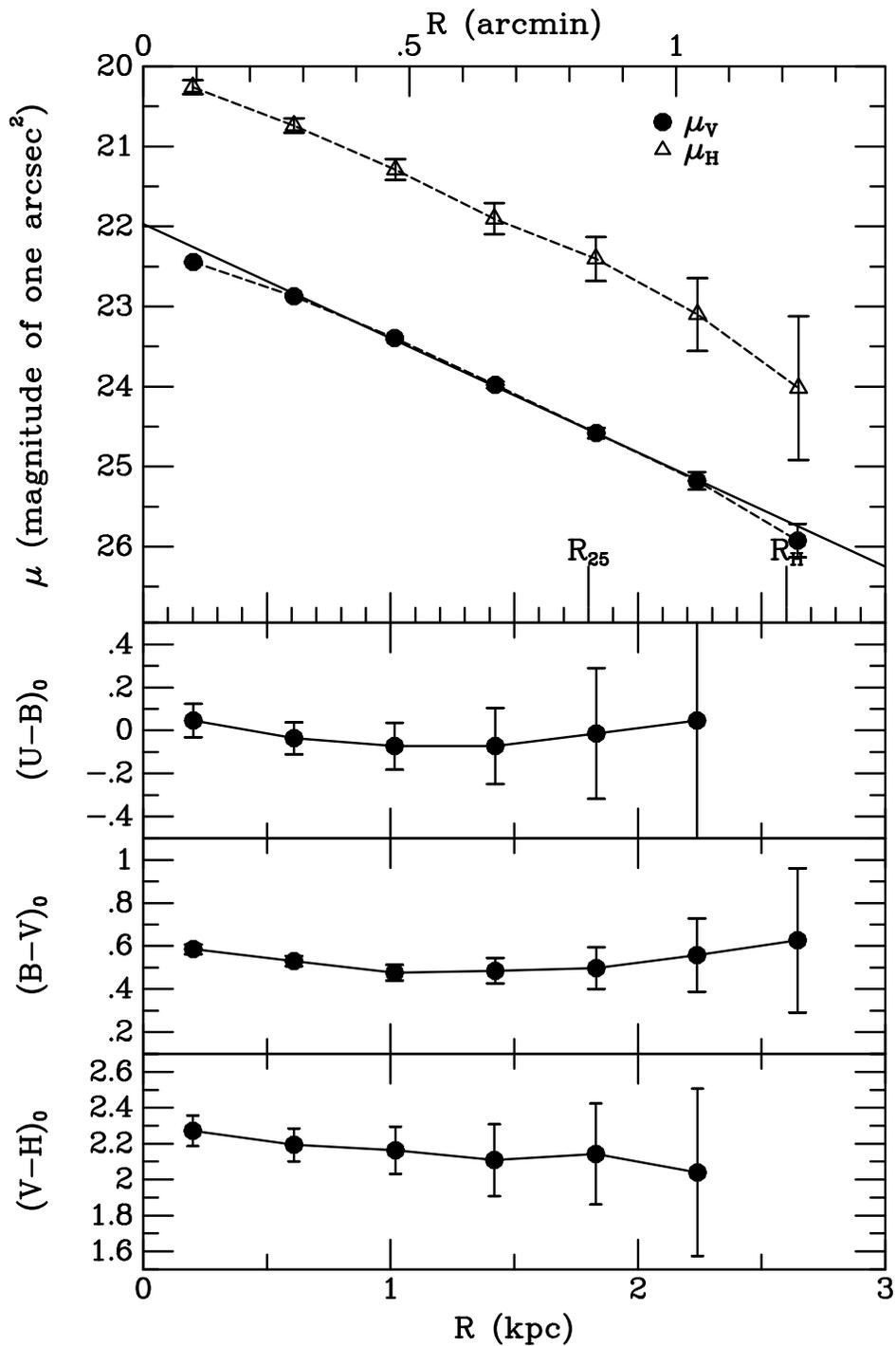}
\caption{UBVH surface photometry of DDO 88.
The photometry is corrected for an internal reddening E(B$-$V) of 0.05 mag
and an external reddening of 0.01 mag.
The exponential fit to
the V-band surface brightness profile is shown as a solid line.
\label{figubvh}}
\end{figure}

UBV and H surface photometry are shown in Figure \ref{figubvh}.  Because the
galaxy is nearly face-on, we used circular apertures increasing in
radius from 11.3\arcsec\ to 79.4\arcsec\ in steps of 11.3\arcsec.  All
photometry has been corrected for reddening using a total
E(B$-$V)$_t$=E(B$-$V)$_f$$+$0.05, where the foreground reddening
E(B$-$V)$_f$ is 0.01 (Burstein \& Heiles 1984). We use the reddening
law of Cardelli \et\ (1989) and A$_V$/E(B$-$V)$=$3.1.

\placefigure{figubvh}

Integrated magnitudes and colors are given in Table \ref{tabglobal}.  We
measured an integrated B magnitude of 14.00$\pm$0.02. This can be
compared to a value of 13.97 given by de Vaucouleurs \et\ (1991) and
13.95 measured by Knezek \et\ (1999), both corrected for extinction
using our values.  Thus, M$_{B,0}$ is $-15.34\pm0.02$ to a radius of
79\arcsec.  The integrated (U$-$B)$_0$ is $-0.03\pm0.09$ and (B$-$V)$_0$
is $0.52\pm0.02$. These are within 0.05 mag and 0.03 mag, respectively,
of the values given by de Vaucouleurs \et\ (1991).  The integrated
(V$-$H)$_0$ is $2.30\pm0.03$, which is similar to NGC 1156 (2.1;
Hunter \& Gallagher 1985), but redder than NGC 4449 (1.9; Hunter, van
Woerden, \& Gallagher 1999) and NGC 2366 (1.3; Hunter, Elmegreen \& van
Woerden 2001) when
measured to comparable radii. The agreement of the integrated colors and
magnitudes with other values in the literature gives us confidence in
our calibration and photometry.  The integrated colors show that DDO 88
lies at the red end of the distribution of a large sample of Im and Sm
galaxies in a UBV color-color plot (see Figure 2 of Hunter 1997).

\placetable{tabglobal}

From $\mu_{B,0}$ as a function of radius, we measured R$_{25}$ to be
0.83\arcmin\ ($=$1.8 kpc).  Our radius is 26\% smaller than that given
by de Vaucouleurs \et\ (1991), but 11\% larger than the radius
measured by Knezek \et\ (1999).  The Holmberg radius, R$_H$,
originally defined to a photographic surface brightness, is measured
at an equivalent B-band surface brightness $\mu_B = 26.7 -
0.149(B-V)$. For a (B$-$V)$_0$ of 0.5, the Holmberg radius is
determined at a $\mu_{B,0}$ of 26.6 magnitudes arcsec$^{-2}$.  We are
able to measure $\mu_{B,0}$ only just to the Holmberg radius.  We find
that R$_H$ is 1.2\arcmin\ ($=$2.6 kpc).

We fit a line to $\mu_{V,0}$ in the inner 17\arcsec--62\arcsec, and the
fit is shown as the solid line in Figure \ref{figubvh}.  DDO 88 is fit
well with an exponential disk profile having a central V-band surface
brightness $\mu_0$ of 21.97$\pm$0.04 magnitudes arcsec$^{-2}$.  If $\mu
= \mu_0 + r/\alpha$, $\alpha_V$ is $0.326\pm0.005$\arcmin\ or
$0.70\pm0.01$ kpc.  Our value is low compared to the $\alpha_B$ and
$\alpha_R$ of 50\arcsec\ and 48\arcsec\ measured by Knezek \et\ (1999).
De Jong (1996) has determined disk characteristics of a large sample of
galaxies spanning a wide range in type.  DDO 88 has a central surface
brightness, corrected to the B passband, that is a little high and a
disk scale length that is a little low compared to what de Jong finds
for late-type galaxies.

As in most irregulars, colors in DDO 88 are quite constant.  Figure
\ref{figubvh} indicates that DDO 88 does not exhibit any major color
gradient with radius. The slopes of the V and H surface photometry plots
are $0.70\pm0.01$ and $0.78\pm0.02$, respectively.  In addition
two-dimensional ratio images confirm what is seen in the
azimuthally-averaged colors: that there are no regions other than the
central cluster and star-forming regions associated with \HII\ regions
that are significantly different in color from the rest of the galaxy.

\begin{figure}
\epsscale{.7}
\plotone{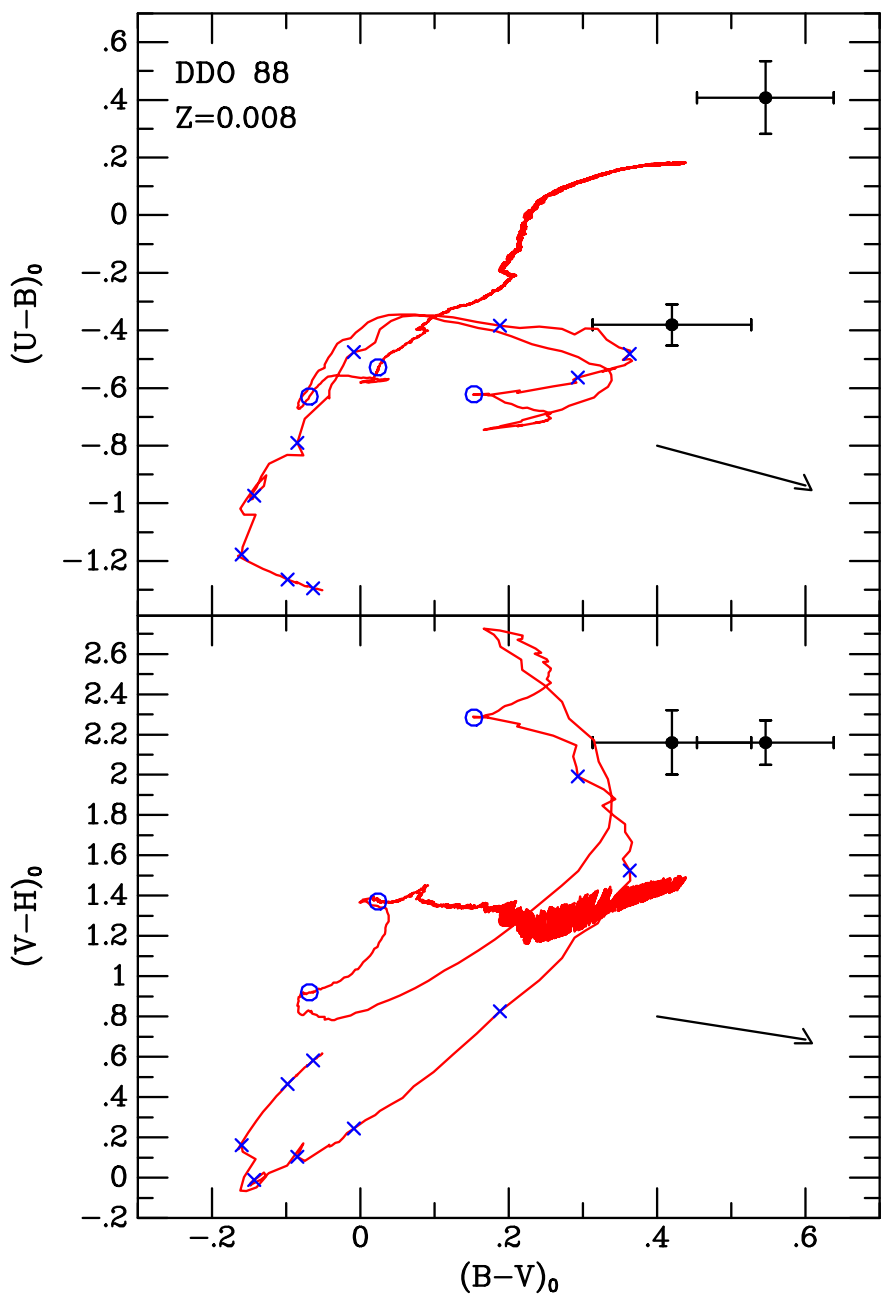}
\caption{Cluster and \protect\HII\ region 2 colors plotted with Leitherer
\protect\et\ (1999) cluster evolutionary tracks for Z=0.008. The arrows are reddening
lines; the photometry has been corrected for modest reddening
(E(B$-$V) = 0.06 mag). The cluster is the redder of the two objects. The
crosses mark 1-9 Myrs in 1 Myr steps; the circles mark 10, 20, and 30
Myr; the tracks end at 1 Gyr.
\label{figtracks}}
\end{figure}

The integrated colors of the central cluster are (U$-$B)$_0 = 0.41$ and
(B$-$V)$_0 = 0.55$, which are red. This could indicate either that the
cluster is old, or heavily reddened. Photometry of the nearby \HII\
region (region 2), however, shows no excessive reddening ((U$-$B)$_0 =
-.30$ and (B$-$V)$_0 = 0.42$), which is confirmed by the
spectroscopy. Reddening in the center of the galaxy near the cluster is
therefore modest, so it is likely that the cluster's colors are due to
age. We have used the Leitherer \et\ (1999) cluster evolutionary models
to compute evolutionary tracks for both the cluster and the nearby \HII\
region (region 2) (Figure~\ref{figtracks}). The \HII\ region colors are
consistent with an age of 7--10 Myr, as expected. The red central
cluster colors agree with an age of 1--3 Gyr. This will be discussed
further in Section~\ref{holeformation}.

\placefigure{figtracks}

\subsection{Star-Forming Regions}

Figure 1b of Knezek \et\ (1999) displays a nice color-composite image
that shows their \ha\ image of DDO 88 on broad-band B and R images.  One
can see a modest number of \HII\ regions spread over the galaxy. The
furthest distinct \HII\ region is found at a radius of 2.0 kpc from the
center of the galaxy.  This radius is 1.1R$_{25}$ or 0.8R$_H$.  The
\HII\ regions are quite modest in luminosity.  If we take the \ha\
luminosity of the Orion nebula as $10^{37}$ \ergsec, the four brightest
\HII\ regions are only 2--6 Orions in luminosity.

The integrated \ha\ luminosity and inferred star formation rate are
given in Table \ref{tabglobal}. We have corrected the \ha\ fluxes for
reddening assuming an internal E(B$-$V) of 0.1 mag and an external
reddening of 0.01 mag.  The total \ha\ luminosity for the galaxy
represents about 80 times that of the Orion nebula. The star formation
rate (SFR) is determined from the \ha\ luminosity using the formula of
Hunter \& Gallagher (1986) which assumes a Salpeter (1955) stellar
initial mass function from 0.1 M\solar\ to 100 M\solar.  In order to
compare to other galaxies, we normalize the SFR to the size of the
galaxy. Here we use the size as defined by R$_{25}$. The normalized
SFR is quite normal compared to other Im and Sm galaxies (see Figure 7
of Hunter 1997).  At its current rate of consumption, the galaxy can
turn gas into stars for another 19 Gyr if all of the gas (\HI$+$He)
associated with the galaxy can be used. The timescale to run out of
gas becomes even longer if recycling of gas from dying stars is also
considered (Kennicutt, Tamblyn, \& Congdon 1994).  As Knezek \et\
(1999) point out, this galaxy can form stars at this rate for a very
long time yet to come.
We note that the L$_{H\alpha}$ given by Knezek \et\ is a factor
of 5 higher than what we measure here. We have not been able to
determine the cause of this difference.

\begin{figure}
\plotone{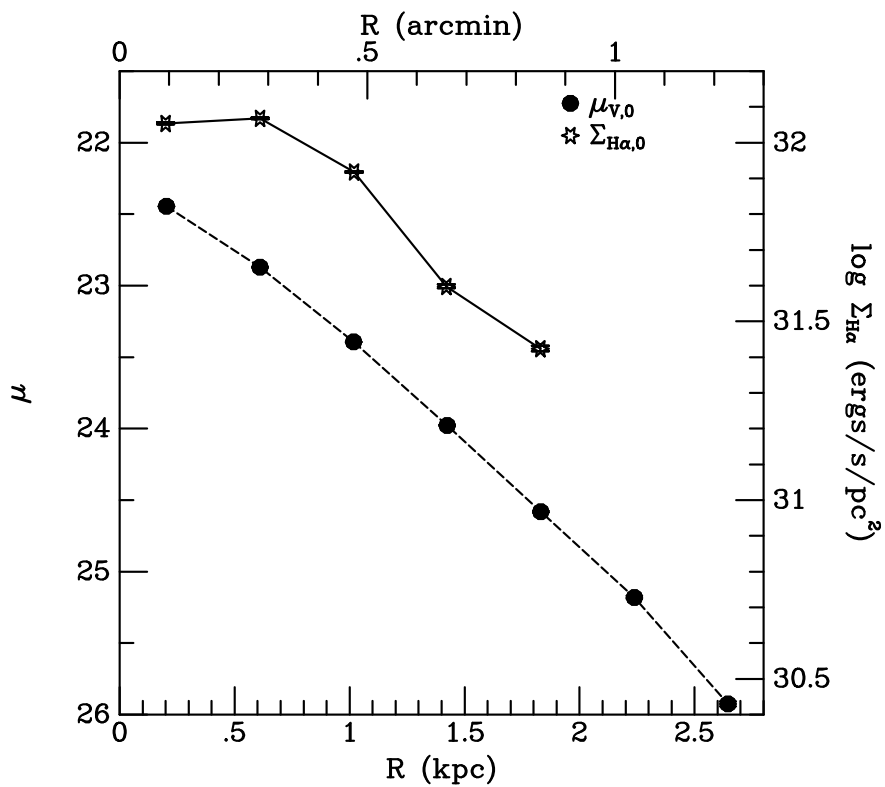}
\caption{Azimuthally-averaged H$\alpha$ surface brightness and
reddening-corrected V-band
surface brightness. The H$\alpha$ luminosity is proportional to the
star formation rate and was corrected for extinction 
using an A$_{H\alpha}$ of 0.28 magnitude.
The scales for $\Sigma_{H\alpha,0}$ and $\mu_V^0$ have been set
so that they cover the same logarithmic interval.
\label{figha}}
\end{figure}

In Figure \ref{figha} we show the azimuthally-averaged \ha\ surface
brightness and compare it to $\mu_{V,0}$.  Outside of the center which
is depressed in \ha\ emission, the \ha\ surface brightness drops at a
rate comparable to that of the starlight.  As in most Im and Sm
galaxies, the current star formation activity ends before the detected
starlight ends.  The outer 36\% of the galaxy by area in which we
detect starlight, as defined by the Holmberg radius, contains no
detectable \HII\ region.
 
\placefigure{figha}

\section{Results: \HI} 

\subsection{\HI\ Morphology}

\begin{figure}
\centering
\includegraphics[scale=0.6,angle=270]{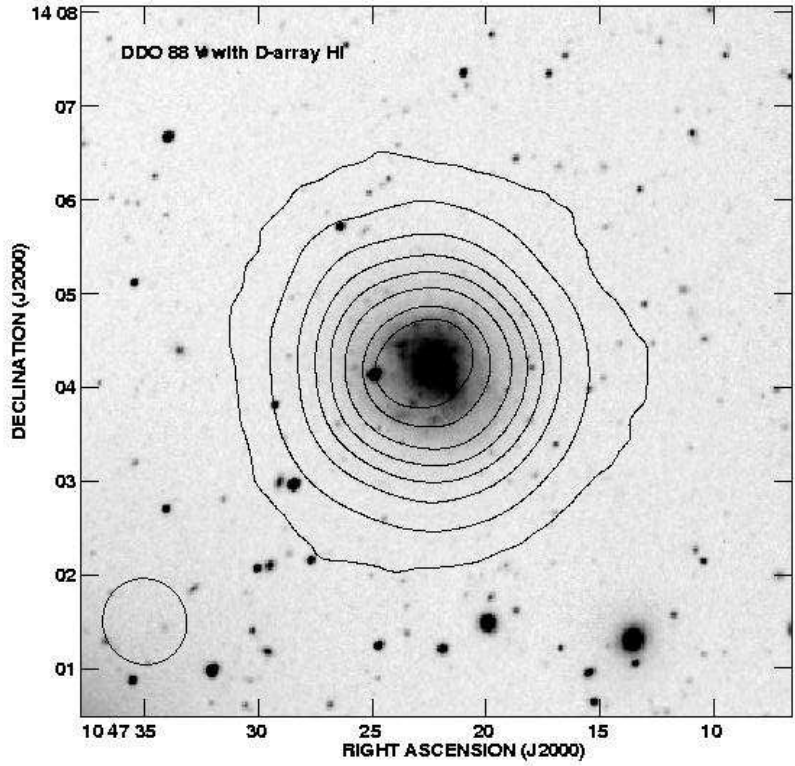}
\caption{V band image with \protect\HI\ contours from the D-array
flux map. Contour levels are at 0.2, 3.7, 11.1, 18.5, 25.9, 33.3,
40.8, and 44.5 $\times\ 10^{19}$ \protect\coldens.
The size of the beam (55.6\protect\arcsec$\times$53.6\protect\arcsec)
is indicated in the lower left
corner.\label{figdonv}}
\end{figure}

Measured \HI\ parameters are shown in Table~\ref{tabhiprop}. In the
high sensitivity D-array integrated flux map the diameter of 
the \HI\ at a column density of $1 
\times 10^{19}$ \coldens\ is 4.1\arcmin\ (8.8 kpc). The total \HI\
mass detected in the system is $7.8\times10^{7}$ \msun. These data
lack the resolution to detect any structure in the \HI\ distribution,
however, and the map shows essentially circular contours of smoothly
increasing density towards the center of the system. The \HI\ contours
on the V band image are shown in Figure \ref{figdonv}.

\placetable{tabhiprop}

\placefigure{figdonv}

\begin{figure}
\plotone{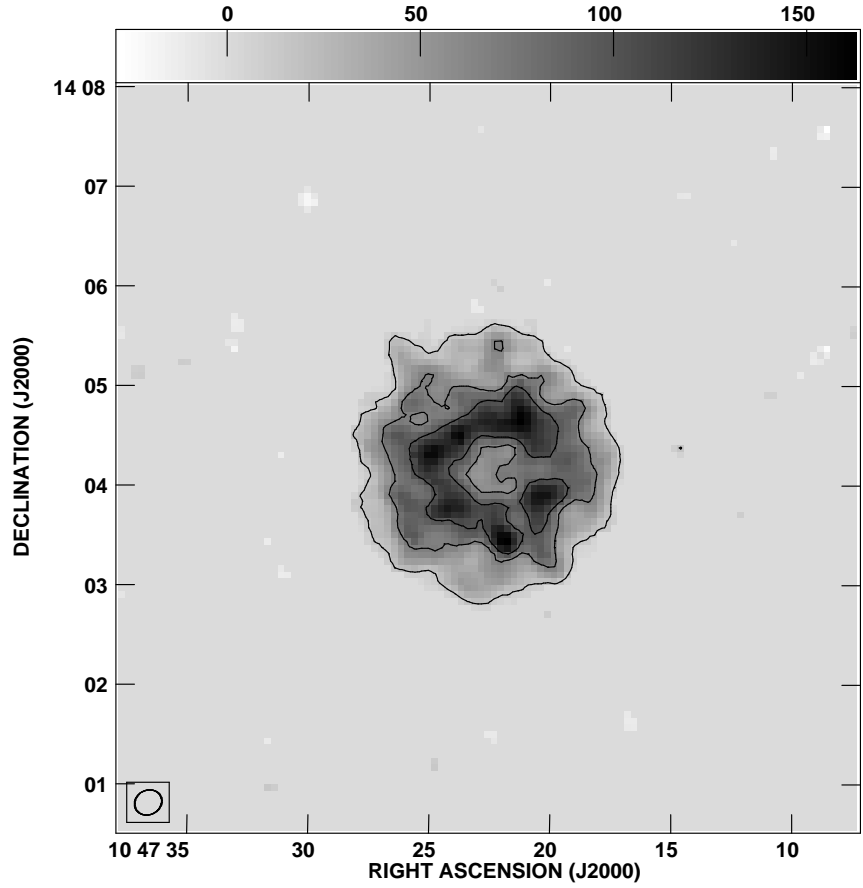}
\caption{Integrated \protect\HI\ flux map from the combined \protect\cd-array
configuration. Contour levels are at 0.7, 3, and 4.5 $\times\ 10^{20}$
\coldens. The size of the beam
(16.85\protect\arcsec$\times$14.61\protect\arcsec) 
is indicated in the lower left corner.
\label{figcdm0}}
\end{figure}
\begin{figure}
\centering
\includegraphics[scale=0.6,angle=270]{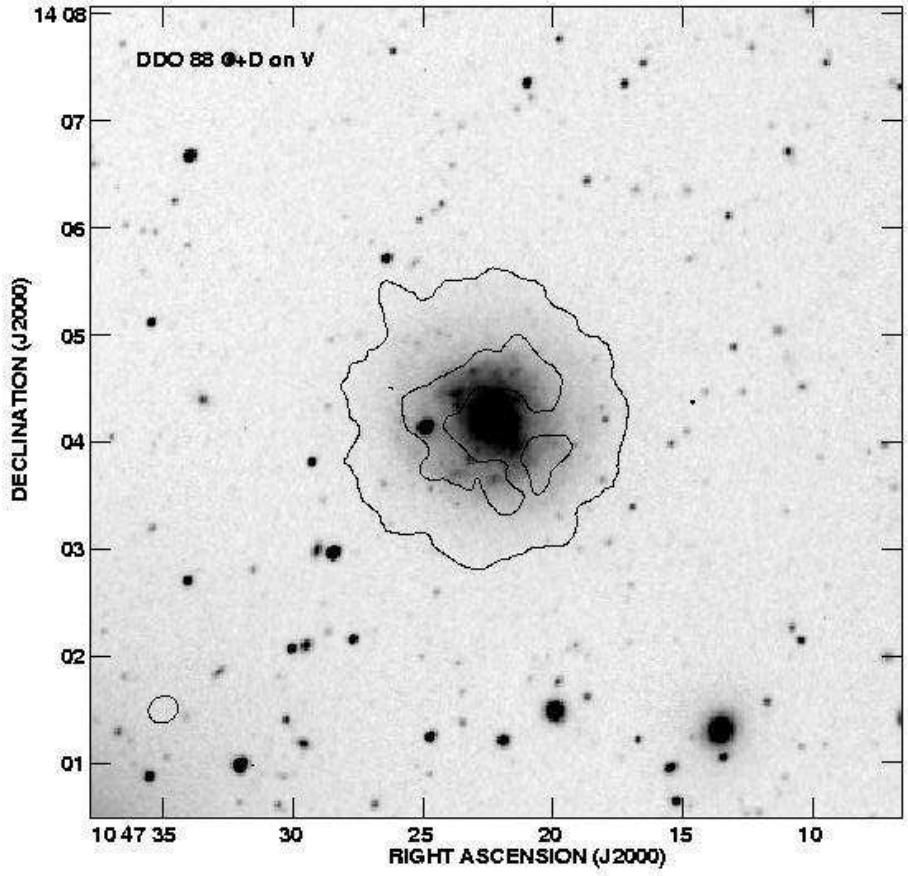}
\caption{V band image with \HI\ contours from the \cd-array flux
map. The contours are at 0.7 and 4.5 $\times\ 10^{20}$
\protect\coldens. The size of the beam (16.85\protect\arcsec $\times$
14.61\protect\arcsec) is indicated in the lower left corner.
\label{figcdonv}}
\end{figure}

The \cd\ array flux map (Figure \ref{figcdm0}), with its decreased
sensitivity to low-level emission, can only be measured out to
$\sim 1 \times 10^{20}$ \coldens. The diameter at the $1 \times
10^{20}$ \coldens\ level, 2.8\arcmin\ (5.9 kpc), is 32\% smaller than
that detected in the D-array data, as it misses the low-level outer
envelope. The higher resolution in this data set is better able to
resolve the structure in the higher density \HI\ gas, however.  The
most striking feature in this flux map is a high density ring in the
\HI\ distribution. Defining the \HI\ ring as the region containing
emission at levels greater than $4.5\times10^{20}$ \coldens, it has an
outer diameter of 3.4 kpc with an average radial thickness of 0.8 kpc
and and average column density of $5.5\times10^{20}$ \coldens. The
ring contains $2.3\times10^{7}$ M\solar\ of \HI\ 
which is 30\% of the total \HI\ mass as measured from the D-array
data. This ring surrounds a central depression or hole in the \HI\
which has a diameter of 49\arcsec\ (1.8 kpc), and apparently surrounds
the highest surface brightness region of the optical (V band)
emission, as demonstrated in Figure \ref{figcdonv}. The star cluster
is located near the center of the hole, but is not visible in this image.

\placefigure{figcdm0}
\placefigure{figcdonv}

\begin{figure}
\plotone{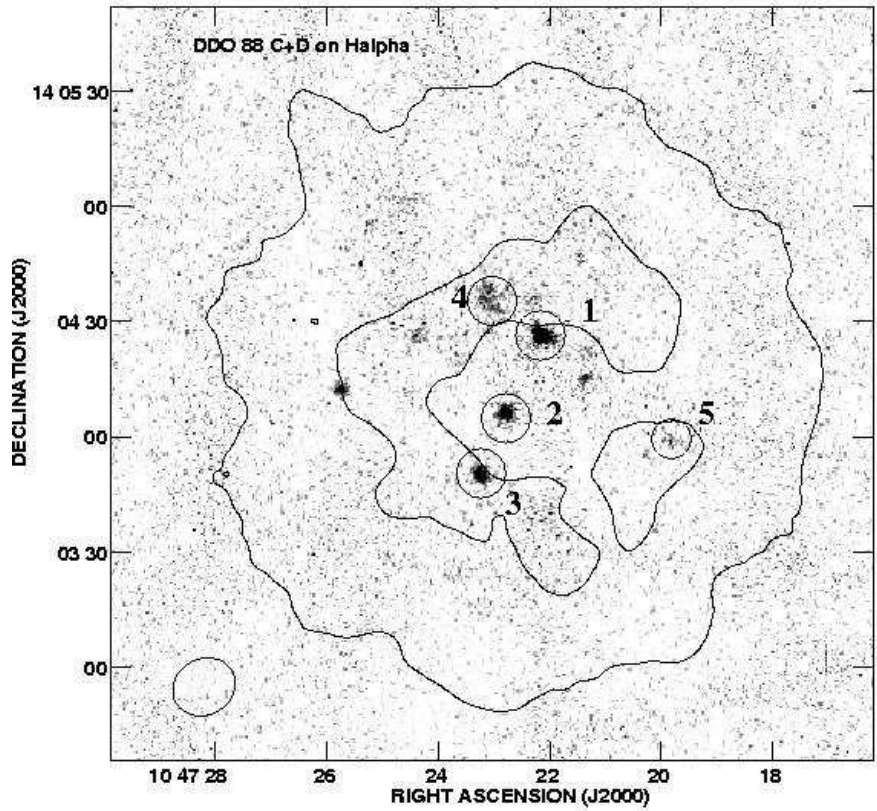}
\caption{\protect\ha\ image with \protect\HI\ contours from the
\protect\cd-array flux map. The contours are at 0.7 and 4.5 $\times\
10^{20}$ \protect\coldens. The size of the beam
(16.85\protect\arcsec$\times$14.6\protect1\arcsec) is indicated in the
lower left corner. The five \protect\HII\ regions listed in Table
\protect\ref{tabhii} are labelled. Note that the red cluster is not
visible in this image.
\label{cdonha}}
\end{figure}

There are several discrete clumps of \HI\ embedded in the ring which are
not resolved.  The average column density in these knots is about
$6\times10^{20}$ \coldens, with peak column densities ranging from
6.6--$7.3\times10^{20}$ \coldens. The largest knot, in the south part of
the ring, contains about $2\times10^6$ \msun. These knots do not appear
to coincide with any noticeable optical features in either the V or \ha\
images. Instead, two of the bright \ha\ regions (regions 1 and 3) are
located on either inside edge of the ring in the southeast and the
northwest, and the third (region 2) is inside the hole in the \HI\ ring
(Figure \ref{cdonha}).

\placefigure{cdonha}

\subsubsection{Surface density profile} 

\begin{figure}
\plotone{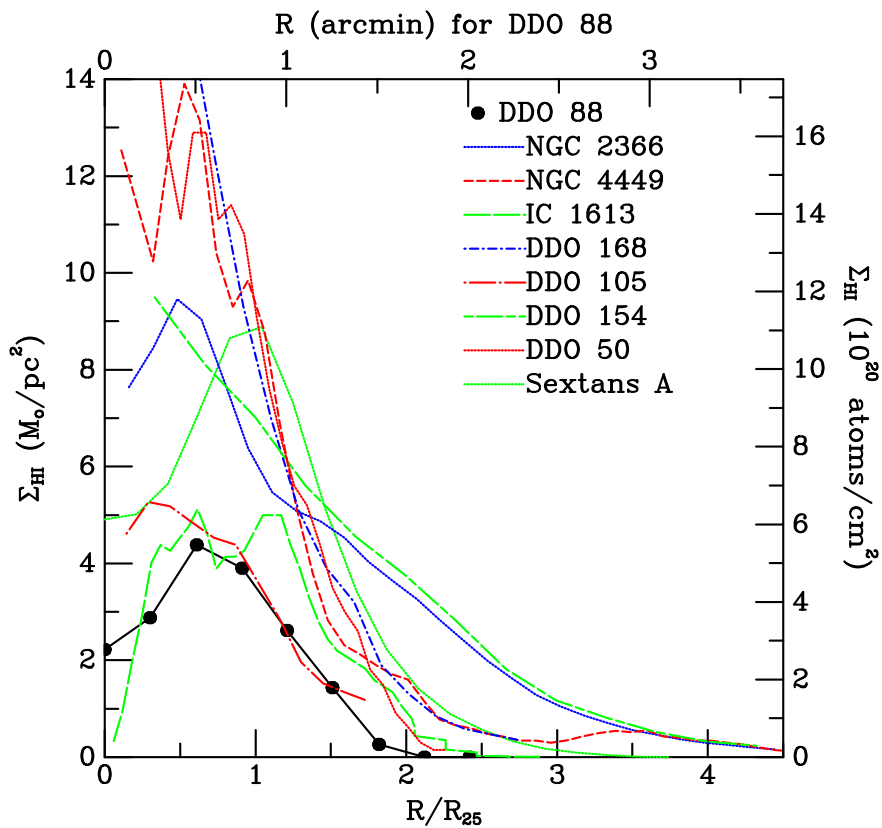}
\caption{Surface density of \protect\HI\ in DDO 88. This is computed
in 15\protect\arcsec-wide circular rings. 
We used the position angle (215\protect\arcdeg) and inclination
(28\protect\arcdeg) 
determined from the \protect\HI\ kinematics.
For comparison we show the \protect\HI\ surface density profiles of a sample
of other Im galaxies (NGC 2366: Hunter, Elmegreen, \& van Woerden
2001; NGC 4449: Hunter, van Woerden, \& Gallagher 1999;
IC 1613: Wilcots 2001;
DDO 105 and DDO 168: Broeils 1992;
DDO 154: Carignan \& Beaulieu 1989;
DDO 50: Puche \protect\et\ 1992; and Sextans A: Wilcots \& Hunter 2002).
\label{figsurden}}
\end{figure}

In Figure \ref{figsurden} we show the azimuthally-averaged surface
density of the \HI\ gas in DDO 88.  We have integrated the \HI\ in
15\arcsec\ radial steps using a position angle of 215\arcdeg\ and
inclination of 28\arcdeg\ as determined from the \HI\ kinematics.  Along
with DDO 88, we plot other irregular galaxies for comparison.  From the
Figure we see that, outside the center ($>$30\arcsec), the gas surface
density drops off fairly smoothly.  The rate at which the \HI\ drops off
with radius is similar to that of other irregulars such as DDO 105 and
IC 1613. However, the drop-off is somewhat shallower than DDO 50's,
Sextans A's, and DDO 168's, but steeper than NGC 2366's and DDO 154's.
In all, the azimuthally-averaged drop-off of the \HI\ surface density
with radius appears to be well within the range of what is observed in
other irregular galaxies.  A central depression in the \HI\ surface
density is seen in other galaxies as well, such as IC 1613, Sextans A,
NGC 2366, NGC 4449, and possibly DDO 105.

\placefigure{figsurden}

The extent of the \HI\ relative to the optical as measured by R$_{25}$
is a little low compared to the collection of irregulars in Figure
\ref{figsurden}, but is nevertheless the same as that of DDO 105 and
cannot be considered too unusual.  In addition the ratio of
R$_{HI}$/R$_H$ in DDO 88 is 1.7 where R$_{HI}$ is measured to a column
density of $1\times10^{19}$ \coldens.  Figure 13 of Hunter (1997)
collects R$_{HI}$/R$_H$ for various irregulars from data in the
literature. DDO 88 is seen to lie at the peak in the number
distribution.

The overall level of the \HI\ surface density is low compared to that
seen in other Im galaxies in Figure \ref{figsurden}; DDO 88 is seen at
the low end of the range in peak $\Sigma_{HI}$.  This may explain why
the star formation activity in this galaxy is so modest, although
generally there is no correlation between maximum gas surface density
and integrated star formation activity in irregulars (Hunter,
Elmegreen, \& Baker 1998).  Furthermore, it is interesting and
puzzling that we see \HII\ regions in DDO 88 in the central part of
the galaxy where the \HI\ surface density is even lower.  Generally in
Im galaxies the star-forming regions are found where the gas surface
density is locally higher than the azimuthally-averaged surface
density (van der Hulst \et\ 1993; van Zee \et\ 1997; Meurer \et\ 1998;
Hunter, Elmegreen, \& van Woerden 2001).

\subsection{Kinematics}

\begin{figure}
\centering
\includegraphics[scale=.75,angle=270]{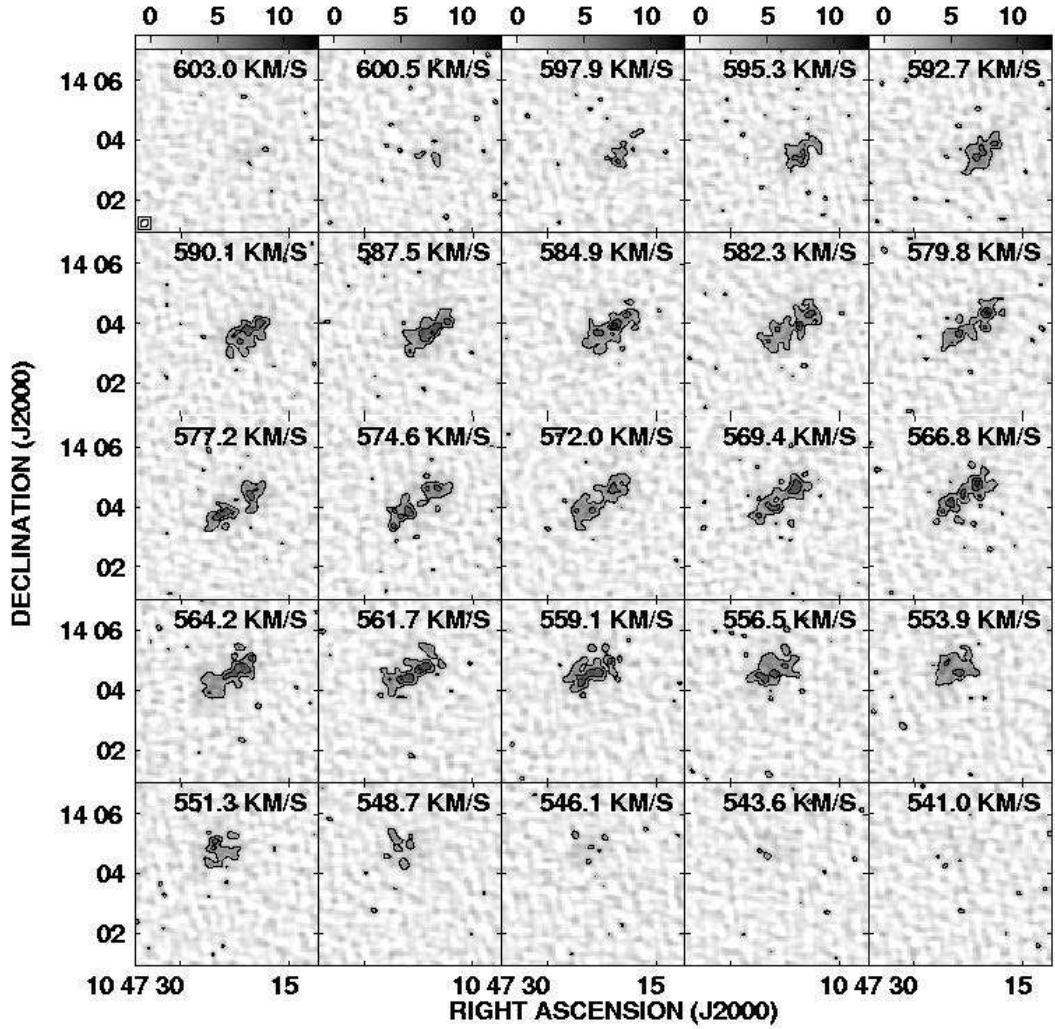}
\caption{\protect\HI\ channel maps from the \protect\cd\ combined-array
cube. Contour levels are at 3 ($3\sigma$), 6, and 9 mJy/B. The cross
indicates the location of the high dispersion knot.
\label{figchanmaps}}
\end{figure}

The \cd-array channel maps (Figure \ref{figchanmaps}) show \HI\
emission between 541 \kms\ and 606 \kms.  One can see mostly ordered,
largely solid body rotation, with ``clumps'' of \HI\ embedded in low
level gas apparent in each channel. Many dwarf galaxies exhibit solid
body rotation over much of their areas (Swaters 1999), and DDO 88
would seem to be no exception.

\placefigure{figchanmaps}

\subsubsection{Rotation curve}

\begin{figure}
\plotone{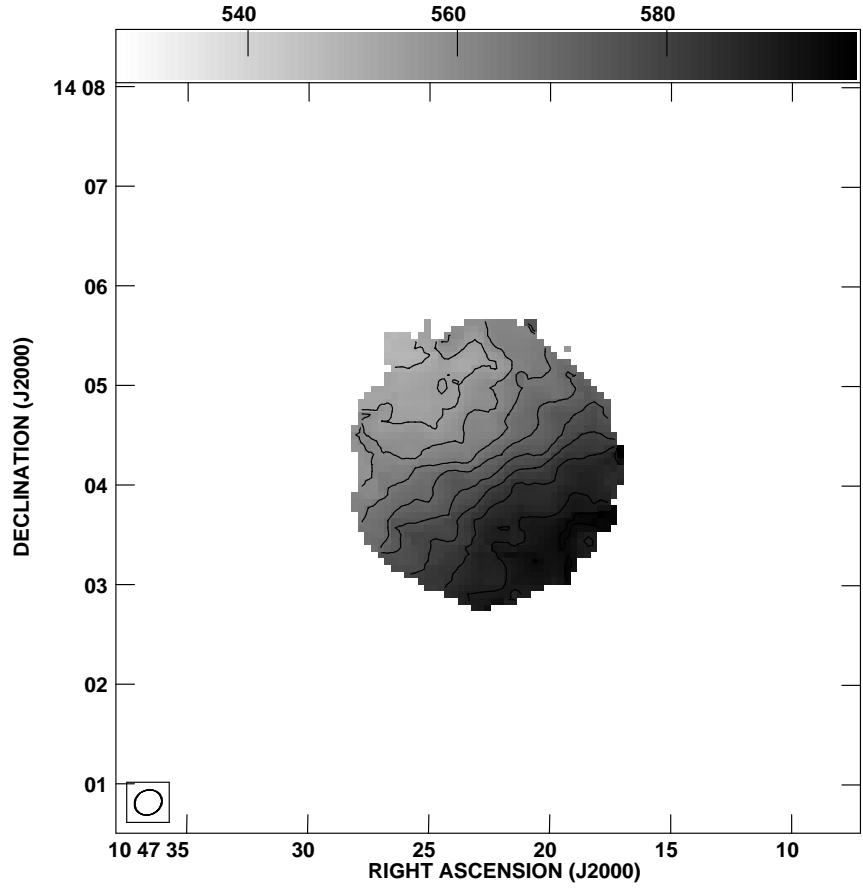}
\caption{Velocity field of \protect\HI\ from the integrated \protect\cd-array
data. Contour levels are every 5 \protect\kms\ from 550--595 \protect\kms.
\label{figcdm1}}
\end{figure}

The velocity field of DDO 88 (Figure \ref{figcdm1}) clearly shows
ordered rotation, so we 
have fit a rotation curve to the velocity field.  We began by fitting
the first moment map of the \cd\ array data with a Brandt function
and allowing all parameters to vary.  The fit was reasonable, and from
this we fixed the center at 10$^h$ 47$^m$ 22.52$^s$ $\pm$ 0.02$^s$,
14\arcdeg\ 04\arcmin\ 10.5\arcsec\ $\pm$ 0.3\arcsec.  With the center
fixed, we fit a solid body rotation law to the inner 40\arcsec\ radius
and determined a systemic velocity V$_{sys}$ of 573.2$\pm$0.1 \kms.
Then fixing the center and central velocity, we fit concentric rings
10\arcsec\ wide and in 10\arcsec\ steps. First, we allowed the
position angle, inclination, and radial velocity to vary. We found
that the position angle and inclination were fairly constant with
radius beyond the inner 20\arcsec\ radius, so we determined an average
position angle of 215.4\arcdeg$\pm$3.5\arcdeg\ and an average
inclination of 28\arcdeg$\pm$8\arcdeg.  With all but the rotation
velocity fixed, we then reran the fits to the velocity field in
concentric annuli.  We also fit the velocity field of the D-array data
alone.  The D-array data are of lower resolution but sensitive to more
extended \HI\ emission.  To fit the D-array data we assumed and fixed
the center, V$_{sys}$, PA, and $i$ determined from the \cd\ array
data and solved for the rotation velocity in annuli of 25\arcsec\
width.

\placefigure{figcdm1}

\begin{figure}
\plotone{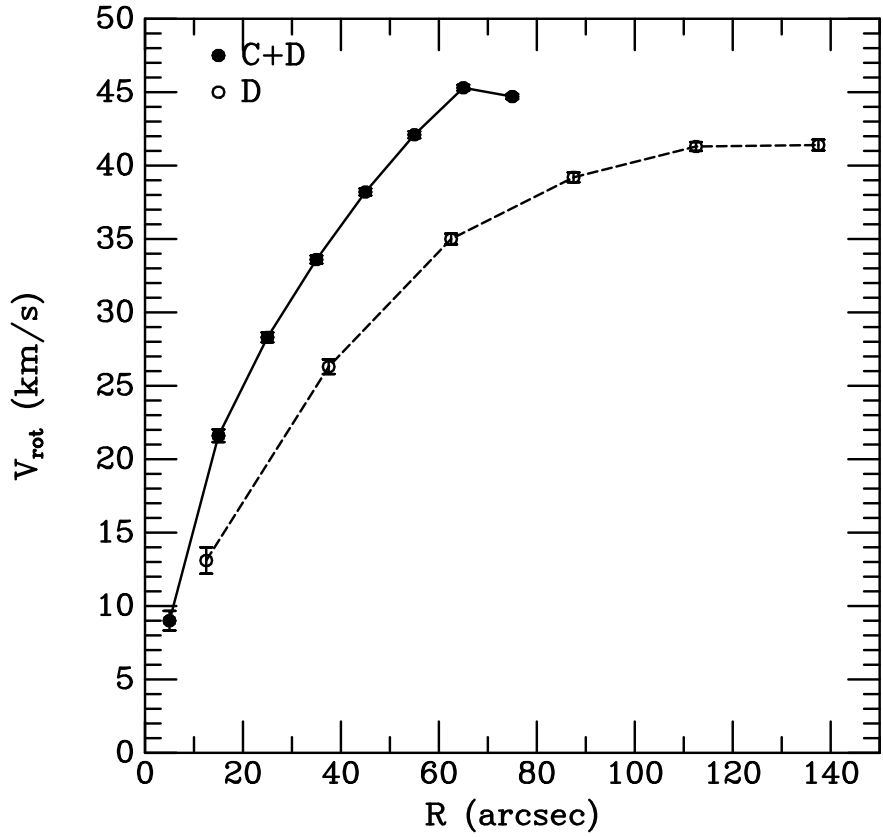}
\caption{Rotation speed as a function of radius, determined from the
\protect\HI\ velocity field. The rotation curves are from the
\protect\cd\ array data that had a beam-size of
16.85\protect\arcsec$\times$14.6\protect\arcsec\ and from the D-array
data alone that had a beam-size of
55.6\protect\arcsec$\times$53.6\protect\arcsec.
\label{figrot}}
\end{figure}

The final rotation curves for the \cd\ array data and for the D-array
data alone are shown in Figure \ref{figrot}.  One can see that the
rotation curve from the \cd\ array data rises steeply to a radius of
about 20\arcsec\ and then rises more shallowly, possibly leveling off
at a radius of about 65\arcsec.  The maximum rotation speed is 45
\kms\ at a radius of 65\arcsec.  The rotation curve from the D-array
data alone rises more gradually and appears to level off around a
radius of 90\arcsec\ with a maximum rotation speed of 41 \kms.  The
shallower rise in the D-array rotation curve is undoubtedly due to the
large beam size (55\arcsec) compared to that of the \cd\ array data
(beam size of 17\arcsec).

\placefigure{figrot}

The rotation curve and maximum rotation speed observed in DDO 88 are
normal for its luminosity. Figure 1 of Hunter, Hunsberger, \& Roye
(2000) plots the maximum rotation speed against M$_B$ for a sample of
Im and Blue Compact Dwarf galaxies. With an M$_B$ of $-15.3$ and a
maximum rotation speed of 45 \kms, DDO 88 lies very close to the
ridgeline of the sample.  Figure 4:5 of Swaters (1999) shows that the
shape of the rotation curve is also fairly normal.

\subsubsection{Velocity dispersion}

\begin{figure}
\centering
\includegraphics[scale=0.6,angle=270]{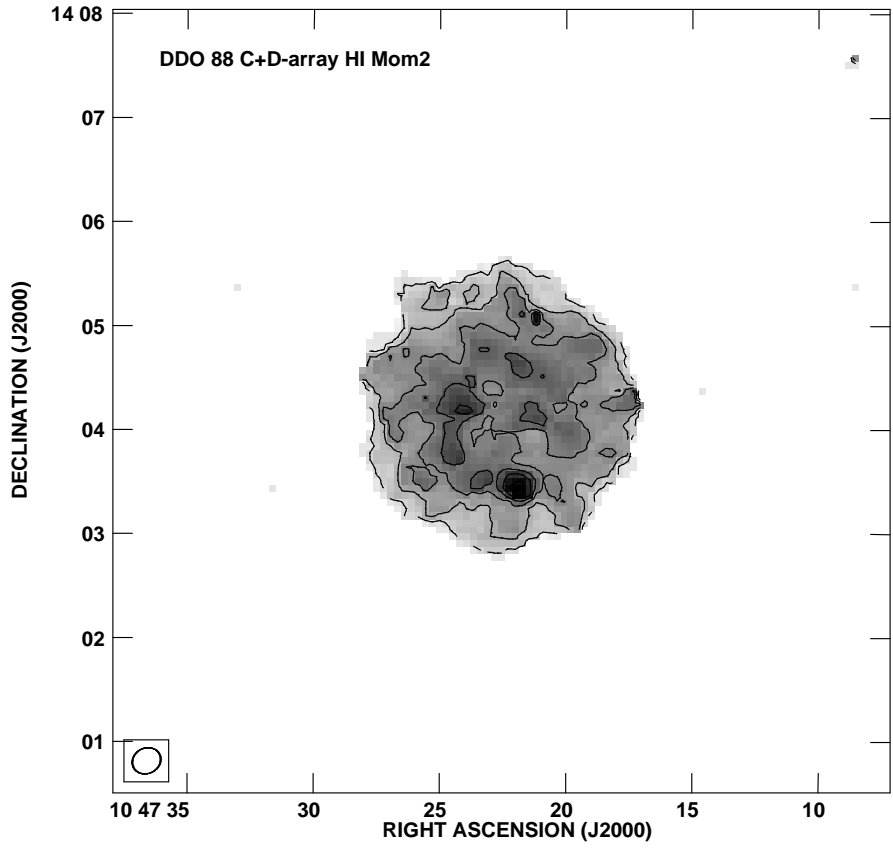}
\caption{Second moment (velocity dispersion) map from the combined
\protect\cd-array data. Contour levels are 2, 4, 6, 8, 10, and 12
\protect\kms. The beam size
(16.85\protect\arcsec$\times$14.61\protect\arcsec) is indicated in the
lower left corner. Note the high dispersion knot in the south part of
the galaxy.
\label{figcdm2}}
\end{figure}

\begin{figure}
\plotone{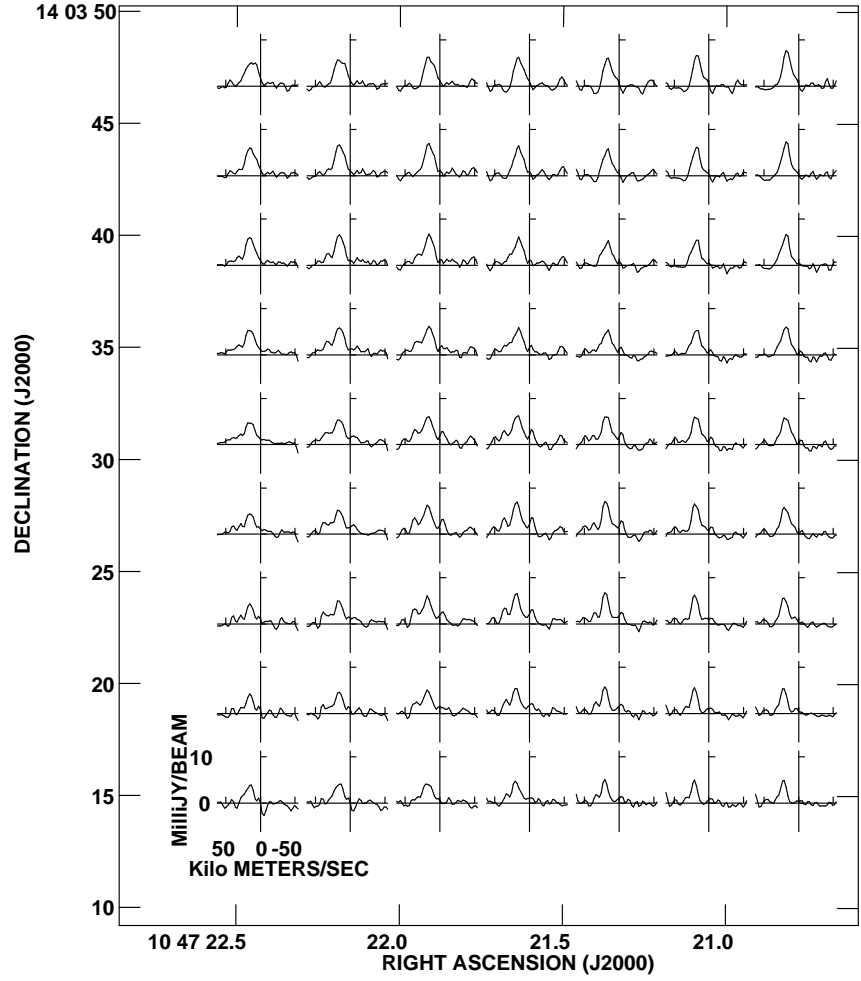}
\caption{Spectra of the high dispersion knot. Each spectrum is from
one pixel through the cube. The beam is approximately $4 \times 4$
pixels. 
\label{figknota}}
\end{figure}

The average velocity dispersion in the second moment map is on the
order of 5 \kms. On average, the gas dispersions are slightly higher
in the area of the ring: 7--9 \kms\ vs.\ 4--5 \kms\ in the outer
regions. There is one notable higher dispersion knot (up to 12 \kms)
located in the southwest (Figure \ref{figcdm2}). This high dispersion
region is associated with the brightest \HI\ knot in the ring, but
interestingly, it is not associated with any significant optical
feature in either the V or \ha\ images. We have fit Gaussians to the
spectra of pixels within the knot; the spectra are shown in
Figure~\ref{figknota}. Most of the spectra, particularly those located
within roughly a beam-width of the center of the knot, exhibit a
central peak with two smaller peaks on either side. Our gaussian fits
indicate that these side components occur at approximately $\pm
20$ \kms\ of the central components. The central components have
amplitudes of about 6 mJy/B, and the side components are roughly half
that. The width of the central components average about 13 \kms, while
the side components average 8 \kms.

\placefigure{figcdm2}

\placefigure{figknota}

The presence of three components could be interpreted as a primary
central cloud of gas that contains a shell of gas that is expanding at
20 \kms. The presence of an expanding shell at that velocity wouldn't
be unusual in a star-forming region, nor should it require large
numbers of stars to form such a shell. However, we would expect to see
\ha\ emission associated with this kind of star-forming event, which
we clearly don't, so this is puzzling. Because we are unable to distinguish
whether the shell is expanding or contracting, perhaps instead we
are seeing the collapse of what will become a star-formation region in
the future.

\section{Discussion}

\subsection{Comparison to other holes}

The 3.4 kpc hole in the center of the integrated \HI\ in DDO 88 is what
intrigued us about this galaxy. Other irregulars do have holes in
their ISM created by the winds and supernova explosions of
concentrations of massive stars or by instabilities in the multi-phase
ISM (Wada, Spaans, \& Kim 2000).  DDO 81 (M$_B=-16.8$) has 6 holes
with diameters greater than 1 kpc, and the largest hole has a diameter
of 1.9 kpc (Walter \& Brinks 1999).  IC 10 (M$_B=-16.5$) has no holes
as large as 1 kpc; all are 100--200 pc in diameter (Wilcots \& Miller
1998).  Holmberg II (M$_B=-16.6$) has 13 giant ($d\geq1$ kpc) holes
and the largest is 1.9 kpc in diameter (Puche \et\ 1992).  Holmberg I
(M$_B=-14.6$) has a large hole just south of the center of
the galaxy that is 1.2 kpc across (Ott \et\ 2001). The tiny 
M81dwA (M$_B\sim-11$) has a central minimum in its \HI\ that is 960 pc
in diameter (Sargent \et\ 1983).  Thus, DDO 88 is not alone in having
such a gas hole.  However, holes this large are rare and may be
especially unusual in a galaxy of this luminosity.

\subsection{The Central \HI\ Minimum}

Here we consider the possibility that the central \HI\ minimum is not
a minimum in the gas, but rather that the \HI\ has been replaced with
molecular gas there.  If this were the case, we can estimate the
amount of \HI\ that is ``missing'' and therefore in the form of H$_2$.
Assuming the average surface density of \HI\ in the center before any
conversion to molecular gas would have been equivalent to that in the
ring, we calculated the total \HI\ mass expected before conversion,
then subtracted out the amount of \HI\ currently in the minimum. The
mass of the ``missing'' gas is then $3\times10^6$ M\solar.  This mass would
correspond to $\sim$10 giant molecular clouds (GMC) in the Milky Way
where a typical GMC has a mass of $4\times10^5$ M\solar\ (H$_2+$He)
(Scoville \& Sanders 1987).  To see if this is possible, we consider
the consequences of such a large molecular complex: the ratio of
molecular-to-atomic material, embedded star formation, and optical
extinction.

Molecular clouds in irregular galaxies are usually part of larger \HI\
complexes (Rubio \et\ 1991, Ohta \et\ 1993).  In the irregular NGC
4449, Hunter, Walker, \& Wilcots (2000) found M$_{H_2}$/M$_{HI}$ is of
order 0.1--1.8 in the star-forming regions of the galaxy.  In DDO 88,
if the central region is filled with $3\times10^6$ M\solar\ of molecular
gas, for the measured 
\HI\ mass of $5.5\times10^6$ M\solar\ in the center of the ring, the ratio of
M$_{H_2}$/M$_{HI}$ is 0.5.  If we consider the \HI\ ring to be part of
the \HI\ complex that hosts the central molecular complex, the ratio
is 0.1. These values are within the range of ratios observed in other
irregulars.

If the central region is filled with a large molecular cloud complex,
we would reasonably expect it to be forming stars unless it is caught
in a very special time.  We do see a few small \HII\ regions there,
but not at a level that would indicate massive amounts of molecular
gas.  However, there could be embedded star formation that is not
optically visible. If so, such star formation would be detectable in
the far-infrared. DDO 88 was not detected by {\it IRAS}, but
observations by {\it IRAS} do provide upper limits on the FIR
fluxes. These are given by Melisse \& Israel (1994) as $<$90 mJy for
the flux at 60 \microns\ and $<$130 mJy for the flux at 100
\microns. Thus, the FIR luminosity L$_{FIR}<3\times10^{40}$ \ergsec,
and L$_{FIR}$/L$_{H\alpha}<46$ and L$_{FIR}$/L$_B<0.3$.  Figure 2 of
Hunter \et\ (1989) shows a histogram of L$_{FIR}$/L$_{H\alpha}$ for a
collection of Im and Sm galaxies detected by {\it IRAS}.  We see that
DDO 88's upper limit of L$_{FIR}$/L$_{H\alpha}$ is normal among
irregular galaxies and, because this is an upper limit for DDO 88, the
real value could be low compared to these other galaxies.
L$_{FIR}$/L$_B$ for DDO 88 is already at the low end for irregulars.
The values of L$_{FIR}$/L$_{H\alpha}$ observed for this sample of Im
galaxies were interpreted by Hunter \et\ as consistent with very
little embedded star formation not detected by \ha. Thus, it is
unlikely that there is extensive star formation optically shrouded in
a central molecular complex in DDO 88.

If the central region is filled with molecular gas, we would also expect
the extinction in this region to be much higher than elsewhere in the
galaxy.  As seen in our Figure \ref{figubvh} and Figure 2a of Knezek \et\
(1999), the UBVR colors in the center of the galaxy are only slightly
redder than most of the rest of the galaxy: U$-$B, B$-$V, and B$-$R are
0.1 mag redder than the bluest annulus in the galaxy. Furthermore, the
surface photometry in V is also only slightly (0.2 mag) dimmed in the
center relative to the projected fit to the rest of the
photometry. Thus, the azimuthally-averaged surface photometry of the
galaxy does not indicate significant extinction in the center of the
galaxy. This is in agreement with the reddening (E(B$-$V)) determined
for \HII\ regions in DDO 88 using long-slit spectroscopy; these are
given in Table~\ref{tabhii}. \HII\ region number 2 in the Table is near
the middle of the \HI\ hole. It has an E(B$-$V) of
0.13$\pm$0.05 mag. Foreground reddening is 0.01 mag, so the reddening internal
to the HII region is 0.12$\pm$0.05 mag. This is quite modest, typical of an
\HII\ region in an Im galaxy (Hunter \& Hoffman 1999), and shows that
the \HI\ hole is not heavily reddened.

Thus, we conclude that the central depression in the \HI\ is unlikely
to be filled with gas in molecular form. Although the M$_{H_2}$/M$_{HI}$
ratio would not be unusual, the lack of star formation and of significant
extinction are not consistent with the presence of a large molecular
complex.

\subsection{Formation of the Hole}\label{holeformation}

If the hole was formed by the action of winds and supernova explosions,
what sort of star formation event would be required?  The \HI\ ring
contains 2$\times10^7$ M\solar\ and the hole has an inner radius of 880
pc.  We do not detect expansion in the ring, but the background
velocity dispersion is 6 \kms.  We will assume that the ring 
attained its final size when its expansion velocity slowed to the
background velocity dispersion.  The models of Chevalier (1974) for
expansion into a uniform medium give the total energy as $E_s =
n_0^{1.12} 10^{54}$ ergs for our ring velocity and radius.

From the surface density profile in Figure \ref{figsurden} we see that
the surface density of the \HI\ surrounding the hole is of order
$5\times10^{20}$ cm$^{-2}$ and we assume this as the pre-hole surface
density.  The thickness of the galaxy is a harder quantity to
estimate.  The diameter of the hole, 1.8 kpc, provides a limit.  With
this thickness, we have $n_0$ is 0.1 cm$^{-3}$.  This is probably a
lower limit, and we will take an $n_0$ of 1 cm$^{-3}$ as an upper
limit. For 0.1 cm$^{-3}$, $E_s$ is $1 \times10^{53}$ ergs, and for 1 cm$^{-3}$
it is 10 times higher.

With an estimate of $E_s$ we then turn to the shell models of McCray
\& Kafatos (1987) to solve for \nstar, the number of stars with mass
greater than 7 M\solar\ that formed the shell.  If $n_0$ is 0.1
cm$^{-3}$, \nstar\ is 1500. For a Salpeter (1955) stellar initial mass
function extending from 0.1 M\solar\ to 100 M\solar, the number of O
stars (18--100 M\solar) would be 390 and the total mass in stars
formed in this event would be $2\times10^5$ M\solar.  If $n_0$ is 1
cm$^{-3}$, these numbers would be 10 times higher.

A sanity check is given by NGC 206, a double OB association in M31
which is sitting in an 800 pc diameter \HI\ hole (Brinks 1981). The mass
in stars of the OB associations that presumably made the hole is
$2\times10^5$ M\solar\ (van den Bergh 1966).  Thus, the hole in DDO
88, which is twice the size, would require 1--10 times more mass in
stars to form. So, this rough estimate seems reasonable.


What would such a star-forming event mean to the galaxy?  We can
estimate the total mass in stars in the galaxy from the V-band
luminosity (M$_V=-15.8$) if we assume an average mass-to-light ratio.
From the models of Larson \& Tinsley (1978), the integrated UBV colors
of the galaxy are consistent with constant star formation over the past
10 Gyr or a burst 500 Myr ago. Even if the central star-forming event
represents a recent burst of star formation, this event is unlikely to
have made the bulk of the stars in the galaxy. Therefore, we assume the
constant star formation model for which the mass-to-light ratio is 1
M\solar/L$_{V,\solar}$. Thus, the total mass of stars in the galaxy is
of order $2\times10^8$ M\solar.  This means the star-forming event
required to make the central hole in the \HI\ would have produced of
order 0.1\%--1\% of the total mass of the galaxy, the range representing
the two limiting $n_0$.  By comparison, the current star-forming
activity is contributing $\sim$0.03\% of the galactic mass in stars.
Therefore, the star-forming event needed to make the \HI\ hole is 4--40
times higher than the current rate and would represent a modest to
vigorous star-forming event in the life of the galaxy.


If the hole was not formed by the action of newly formed massive stars,
how then did it form? Wada, Spaans, \& Kim (2000) have suggested
that holes can form in the ISM of galaxies by the nonlinear
development of the combined thermal and gravitational instabilities
in the disk gas. This does not require the immediate presence of
a starburst. This model is used to explain the multitude of holes
and filaments found in Im galaxies like the LMC and DDO 50.
In their model a spectrum of sizes of holes is formed, with the
typical holes being 500 pc in diameter or less. A few holes of
kpc size are formed, but rarely. Thus, the hole in DDO 88 would
be an extraordinary event even in their model. Furthermore,
there are no other holes in DDO 88---just one big one. This does
not really make sense in the context of the Wada \et\ models.

Thus, we return to the injection of energy by massive stars longer ago
than 0.5 Gyr as the only mechanism for forming the hole in DDO 88 that
makes sense. Is the age of the ring consistent with this?  Could the
red cluster in the center be the remnant of such an event? As we don't
detect expansion in our \HI\ data, the ring has apparently stalled. If
we assume that expansion stopped when the expansion velocity equalled
the dispersion velocity in the gas, this method provides an upper limit
to the time it took to \emph{form} the ring.  Using a dispersion
velocity of 6 \kms\ and a size of 880 pc (the inner radius of the hole), this
gives us an upper limit of approximately 145 Myr during which the hole
and ring were formed. If we use the distance from the center of the
hole to the midpoint of the \HI\ ring (1.28 kpc) as our size instead,
the maximum expansion time becomes 210 Myr. Note that this does not tell
us how long ago the expansion stopped, so in that sense it serves as a
rough lower limit to the age of the ring; with the caveat that the
ring could also have formed somewhat faster than we have calculated here.

Ott \et\ (2001) use a variety of methods to estimate the age of the large
stalled \HI\ shell they observed in the low mass dwarf Holmberg I
(Ho I). This shell, which surrounds a depression in the \HI\ just south
of the center of the galaxy, dominates the appearance of the galaxy. It
extends to approximately half the optical size of the galaxy, similar to
what we see in DDO 88. The size of this shell (center to point of
maximum \HI\ emission) is
850 pc, which is smaller than the 1.28 kpc for DDO 88's ring, but is
similar to the inner radius of the ring surrounding the hole in DDO 88.
Using their shell radius and a 
background dispersion of 9 \kms, they estimate a maximum duration of
expansion of 90 Myr. This agrees with the total age they find from the
central colors using the Leitherer \et\ (1999) models, as well as the
ages of stars observed in the \HI\ shell itself. They also calculate the
age of the shell using a Sedov expansion model (Sedov 1959; Mac Low \&
McCray 1988; Ehlerov\'{a} \et\ 1997) for the time prior to ``break
out,'' when the shell expands out of the disk of the galaxy as defined
by the scale height. After this, the expansion enters a ``snowplow''
phase (McCray \& Kafatos 1987). From this combined shell evolution
model, they again come up with an age estimate of approximately 100 Myr.

We also tried applying this shell evolution model to DDO 88. However,
there are large uncertainties in calculating or estimating several of
the quantities that are used in these equations (equations 7 and 8 in
Ott \et\ 2001). Among these are an estimate of the scale height, which
is used as the ``critical radius'' at which breakout occurs and the time
until break out, which depends on this radius as $R_c^{5/3}$.  Other
parameters used to calculate this ``critical time'' include the number
of supernovae per Myr, $\dot N$, which depends on the energy required to
create the shell and the average energy per supernova, the particle
volume density $n_0$, and the mean molecular mass $\mu$. Both the
critical time and the critical radius are then used to estimate the
duration of the snowplow phase, with the result depending on $t_c
R_c^{-4}$.

Depending on the method we use, we find a scale height for DDO 88 of
between 100--300 pc. For $\dot N$, if we assume (as Ott \et\ did)
$10^{51}$ ergs per supernova, and that they all go off within 40 Myr,
then the number of supernova per Myr = $E_s/10^{51}$/40. For our $E_s =
10^{53}$--$10^{54}$ ergs (corresponding to $n_0 = 0.1--1.0$ cm$^{-3}$)
this is 100-1000 supernovae, so $\dot N = 2.5$--25 Myr$^{-1}$. As we
have used the radius of the hole (not the midpoint of the ring) to
calculate $E_s$, we continue to do so. Thus, for a radius of 880 pc, we
find a ring age of 0.5--6.5 Gyr. If instead we use the midpoint of the
ring as our ring size, we find a range of 1.4--20 Gyr. Clearly, the
upper bound here is unphysical. The lower bounds are in line with the
central colors in the hole that indicate no vigorous star formation in
the past 0.5 Gyr. The red central cluster colors agree with an age of
1--3 Gyr, so it is old enough to have formed the hole.

Could the ring survive as long as a Gyr though? The rotation curve indicates
that DDO 88 is undergoing primarily solid-body rotation, especially in
the inner regions where the ring is located (approximately
30--50\arcsec). Thus the ring would not be subject to the shearing
forces produced by differential rotation. But what about dispersion due
to the random motions in the gas? Recall that we have an upper limit for
the expansion time of the hole of about 150--200 Myr, so if the ring is
more than 0.5 Gyr old, it has been stalled for over 300 Myr. The \HI\
ring has a width of 0.8 kpc. With a dispersion velocity of 6 \kms, it
would take 270 Myr for material to travel twice the width of the
ring. This is close to our lowest estimate for the time since the ring
stopped expanding. Thus, it would seem that the ring should have
undergone at least some dispersion, if not total, if it is much older
than 0.5 Gyr. This argues for a younger age for the ring, and against
the ring being as old as the red cluster. 



The next question is whether the cluster could have produced the
required energy. Currently, the cluster has an M$_V = -9.99$. Using the
lower age limit of 1 Gyr, we find, according to the Leitherer \et\
models, it would have had an M$_V = -13.6$ at 10 Myr and a mass of $4
\times 10^5$ \msun.  This is consistent with the 2--20$ \times 10^5$
\msun\ of stars that we estimated would have formed in the event that
made the hole, and is similar to the mass estimate for the association
in NGC 206 (Brinks 1981). This places it in the realm of super star
clusters, so it very well could have produced the \HI\ hole from an
energy standpoint.

\subsection{Propagating Star Formation?}

Star formation regions have been found at the edges of some \HI\ shells
(e.g.\ Brinks 1981, Walter \& Brinks 1999), presumably as a result of
instabilities that begin to occur in the swept-up gas from the shell
formation (McCray \& Kafatos 1987, Elmegreen 1994). Although there is no
evidence of 
increased star formation activity seen from the azimuthally-averaged
\ha\ profile (Figure \ref{figha}), four of the five \ha\ regions are
associated with the \HI\ ring (Figure \ref{cdonha}). Here, however, we
run into an inconsistency if our ring is indeed older than 0.5 Gyr and
has been stalled for over 300 Myr. The \HII\ region colors indicate ages
of 7--10 Myr, so there would have been a considerable delay between the
formation of the ring and the onset of secondary star formation.  Would
star formation continue to occur in the ring so long after it formed?

There is evidence that secondary star formation resulting from the
formation of an \HI\ shell can occur relatively quickly, but how long it
continues is much less clear. Walter \et\
(1998) observed a still-expanding supershell in IC 2574 with an age of
approximately 14 Myr. There are massive \HII\ regions coincident with
the rim of this shell, so presumably secondary star formation is
occurring now. Constellation III in the LMC has an \HI\ hole 1.4 kpc in
diameter that formed about 20 Myr ago, and \HII\ regions with ages of
approximately 5 Myr are observed in the shell (Dolphin \& Hunter 1998),
indicating that it took less than 15 Myr for them to form. 

Looking at an
older, stalled shell, we turn again to the supershell in Ho I with an
age of $\sim 80$--100 Myr (Ott \et\ 2001). Inside the rim, they
observe stars 15--30 Myr old, with younger regions of star formation
located on the rim of the hole, so it appears that secondary star
formation is occurring some 50--60 Myr after the hole began to
form. However, the upper limit on the age of this shell is 90 Myr, so it
has been stalled for less than 10 Myr.

Therefore, observed cases of induced star formation give timescales
for secondary star formation in the shells of order 60 Myr or less. 
This leaves us with the problem that the one stellar remnant that we 
find in the hole is too old, and we cannot identify any younger entity.
At this point we see no way around this inconsistency and cannot clearly
point to what has formed a galaxy-sized hole in the gas of this tiny
galaxy. 

\section{Summary}

DDO 88 appears to be a fairly normal low-luminosity dwarf galaxy in the
optical, with only modest on-going star formation.  There is a slight
asymmetry in the outer low-level V-band isophotes, but this is not
unusual among Im-type galaxies. Its integrated UBVH colors put it at the
red end of the distribution for Im/Sm galaxies, consistent with the
current low level of star formation and suggesting that star formation
has been roughly constant over the lifetime of the galaxy. Any
significant star formation event would have had to occur more than 0.5
Gyr ago.  The azimuthally-averaged surface photometry is well fit by an
exponential disk, and there are no major color gradients detected across
the disk. These features are typical of Im/Sm galaxies.

In \HI, DDO 88's outer regions and global characteristics also appear
quite normal.  The ratio of \HI-to-optical radius is typical of Im and
spiral galaxies.  The azimuthally-averaged \HI\ surface density is a bit
low compared to other irregular galaxies but it drops off at an average
rate.

The normal appearance of the optical and outer regions of the \HI\ gas
in DDO 88 give no hint that the center of the galaxy hosts a large (1.8
kpc diameter) \HI\ depression. The \HI\ surrounding the hole is
distributed in a high column density ring with an outer diameter of 3.4
kpc. The hole is unusually large for a galaxy of this size, and does not
appear to be expanding. If the hole was produced by the energy injection
from winds and supernova explosions of massive stars formed in a
starburst, the star formation event would have produced $2$--$20\times
10^5$ \msun\ of stars, which is 0.1--1\% of the total mass of the
galaxy.

There is a red star cluster near the center of the \HI\
hole. Corrected for reddening, evolutionary models give  an age
of 1--3 Gyr and an initial mass of $4\times 10^5$ \msun. Both the age
and mass are consistent with the cluster being the remnant of a star
formation event old enough and energetic enough to have created the \HI\
hole in the center of the galaxy. If this is the case, it is possible
that the higher density \HI\ ring surrounding the hole is from gas being
swept up and pushed outward by the stellar winds and supernovae that
accompanied the cluster formation.

Age estimates for the \HI\ ring indicate that is it likely older than
0.5 Gyr, which might agree
with the age of the red cluster, but this is uncertain as the ring is
stalled. An upper limit for the time needed to form the ring is
150--200 Myr. Dispersal time for the ring from random motions in the
gas is on the order of 300 Myr however, which would indicate that this
large feature should have mostly disappeared by now if it is indeed
older than 0.5 Gyr. Additionally, there are faint \HII\ regions in and
on the rim of the ring that are 7--10 Myr old. If they are secondary
star formation regions caused by the ring, this also points to a
younger ring age. With conflicting evidence regarding how long ago the
hole and resultant ring formed, we cannot state that the red central
cluster is the causative agent; nor can we rule it out. We have been
unable to identify any other mechanism that could be responsible for a
such a dramatic feature in this small galaxy.

\acknowledgments

The authors would like to
thank Alan Watson for his help in the acquisition and initial reduction
of the IR data and the referee for helpful comments.

Support to DAH for this research came from the Lowell Research
Fund and in part from grants AST-9802193 and AST-0204922 
from the National Science Foundation.

This publication makes use of data products from the Two Micron All Sky
Survey, which is a joint project of the University of Massachusetts and
the Infrared Processing and Analysis Center/California Institute of
Technology, funded by the National Aeronautics and Space Administration
and the National Science Foundation.

This research has made use of the NASA/IPAC Extragalactic Database (NED)
which is operated by the Jet Propulsion Laboratory, California Institute
of Technology, under contract with the National Aeronautics and Space
Administration.

\clearpage

\begin{deluxetable}{lccc}
\tablecaption{VLA Observations
\label{tabhi}}
\tablewidth{0pt}
\tablehead{
\colhead{} 		& \colhead{D-array} 	& \colhead{CS-array}	& \colhead{C$+$D\tablenotemark{a}}
}
\startdata
Observation Date	& 1996 Sept.\ 20 \& 24	& 1997 Sept.\ 19 	& \nodata\\
Time on Source~(min)	& 449 			& 161 			& \nodata\\
Bandwidth~(MHz)		& 1.56 			& 1.56 			& 1.56\\
No.\ of Channels	& 128			& 128			& 128\\
Velocity Resolution~(\kms)	& 2.6 		& 2.6 			& 2.6 \\
Beamsize\tablenotemark{b}~(\arcsec) & $55.6 \times 53.6$
						& $14.5 \times 13.1$
									& $16.8 \times 14.6$\\
Single channel r.m.s.~(mJy/B)	& 0.71 		& 1.26 			& 0.98\\

\enddata
\tablenotetext{a}{Combined data set from both arrays}
\tablenotetext{b}{Natural weighting for D; Uniform weighting with
robust factor of $-1$ for CS and C$+$D}
\end{deluxetable}


\begin{deluxetable}{lccccc}
\tablecaption{Results of HII Region Spectroscopy.
\label{tabhii}}
\tablewidth{0pt}
\tablehead{
\colhead{} & \multicolumn{5}{c}{HII Region} \\
\colhead{Quantity\tablenotemark{a}} & \colhead{1} & \colhead{2} 
& \colhead{3} & \colhead{4} & \colhead{5}
}
\startdata
~RA\tablenotemark{b} (2000)  & 10:47:22.1 & 10:47:22.8 & 10:47:23.2 & 10:47:23.1 & 10:47:19.8 \\
~DEC\tablenotemark{c} (2000) & 14:04:26   & 14:04:06   & 14:03:50   & 14:04:36   & 14:03:59   \\
~H$\alpha$/H$\beta$ & 2.97$\pm$0.55 & 3.27$\pm$0.15 & 3.00$\pm$0.13 & 2.96$\pm$0.22 & 3.35$\pm$0.36 \\
~[OIII]$\lambda$5007/[OIII]$\lambda$4959 & 2.81$\pm$0.08 & 3.23$\pm$0.57 & 2.90$\pm$0.34
                & 2.84$\pm$0.08 & \nodata \\
~[OII]$\lambda$3727/H$\beta$ & 2.87$\pm$0.93 & \nodata & \nodata & \nodata & \nodata \\
~[OIII]$\lambda$5007/H$\beta$ & 1.22$\pm$0.06 & 0.64$\pm$0.04 & 0.60$\pm$0.03
                & 1.79$\pm$0.08 & 0.26$\pm$0.10 \\
~[NII]$\lambda$6584/H$\alpha$ & 0.15$\pm$0.03 & 0.14$\pm$0.02 & 0.16$\pm$0.04 
                & 0.18$\pm$0.05 & 0.15$\pm$0.07 \\
~[SII]$\lambda$6717$+$6731/H$\alpha$ & 0.13$\pm$0.09 & 0.25$\pm$0.07 & 0.27$\pm$0.05
                & 0.53$\pm$0.14 & 0.43$\pm$0.30 \\
~E(B$-$V)\tablenotemark{d} & 0.04$\pm$0.17 & 0.13$\pm$0.05 & 0.05$\pm$0.04 
                & 0.03$\pm$0.08 & 0.16$\pm$0.10 \\
\enddata
\tablenotetext{a}{Line ratios are not corrected for reddening.}
\tablenotetext{b}{RA are in hours, minutes, seconds of time.}
\tablenotetext{c}{DEC are in degrees, arcminutes, arcseconds.}
\tablenotetext{d}{This assumes an electron density of 100 cm$^{-2}$,
a nebular temperature of 10$^4$ K, and the Cardelli et al. (1989) reddening law.}
\end{deluxetable}

\clearpage
 
\begin{deluxetable}{lc}
\tablecaption{Summary of Integrated Properties.
\label{tabglobal}}
\tablewidth{0pt}
\tablehead{
\colhead{Parameter} & \colhead{Value}
}
\startdata
D (Mpc)\tablenotemark{a} \dotfill                                            	&  7.4 \\
E(B$-$V)$_f$\tablenotemark{b} \dotfill                                       	&  0.01 \\
R$_{25}$ (arcmin) \dotfill                                                   	&  0.83 \\
R$_{25}$ (kpc) \dotfill                                                      	&  1.8 \\
R$_H$ (arcmin) \dotfill                                                      	&  1.2 \\
R$_H$ (kpc) \dotfill                                                         	&  2.6 \\
$\mu_0$ (V-band, magnitudes arcsec$^{-2}$) \dotfill                          	& 21.97$\pm$0.04 \\
$\alpha_V$ (kpc) \dotfill							&  0.70$\pm$0.01 \\
$\alpha_H$ (kpc) \dotfill                                                    	&  0.78$\pm$0.02 \\
M$_{B,0}$ (r$=$79\protect\arcsec) \dotfill				     	& $-15.34\pm0.02$ \\
(U$-$B)$_0$ (r$=$79\protect\arcsec) \dotfill                                 	& $-0.03\pm0.09$ \\
(B$-$V)$_0$ (r$=$79\protect\arcsec) \dotfill                                 	& 0.52$\pm$0.02 \\
(V$-$H)$_0$ (r$=$79\protect\arcsec) \dotfill					& 2.30$\pm$0.03 \\
log L$_{H\alpha,0}$ (ergs s$^{-1}$) \dotfill                                 	& 38.89 \\
SFR\tablenotemark{c}~ (M\protect\solar\ yr$^{-1}$) \dotfill                  	& 0.0055 \\
log SFR/area\tablenotemark{c}~ (M\protect\solar\ yr$^{-1}$ kpc$^{-2}$) \dotfill	& $-$3.26 \\
\enddata
\tablenotetext{a}{V$_{GSR}$ = 481 \kms; H$_0$ = 65 \kms\ Mpc$^{-1}$.}
\tablenotetext{b}{E(B$-$V)$_f$ is foreground reddening due to the Milky Way
(Burstein \& Heiles 1984).
For the broad-band stellar photometry in DDO 88, 
we assume an additional internal reddening of 0.05
magnitude;
for the \protect\HII\ regions
we assume an additional internal reddening 0.1 magnitude.}
\tablenotetext{c}{Star formation rate derived from L$_{H\alpha}$
using the formula of Hunter \& Gallagher (1986) that integrates
from 0.1 M\protect\solar\ to 100 M\protect\solar\ with a Salpeter
(1955) stellar initial mass function.
The area is $\pi$R$_{25}^2$.}
\end{deluxetable}

\clearpage

\begin{deluxetable}{lc}
\tablecaption{HI Properties.
\label{tabhiprop}}
\tablewidth{0pt}
\tablehead{
\colhead{Parameter} & \colhead{Value}
}
\startdata
M$_{\rm{HI}}$\tablenotemark{a} \dotfill			& $7.8\times10^7$ \msun\\
R$_{\rm{HI}}$\tablenotemark{a}~ (arcmin) \dotfill	& 2.1\\
R$_{\rm{HI}}$\tablenotemark{a}~ (kpc) \dotfill		& 4.4\\
M$_{\rm{HI}}$\tablenotemark{b}~ in the ring \dotfill	& $2.3\times10^7$ \msun\\
M$_{\rm{HI}}$\tablenotemark{b}~ in the hole \dotfill	& $5.5\times10^6$ \msun\\
R$_{\rm{hole}}$\tablenotemark{b}~ (arcsec)\dotfill	& 25\\
R$_{\rm{hole}}$\tablenotemark{b}~ (kpc)\dotfill		& 0.9\\
Width of ring\tablenotemark{b}~ (arcmin) \dotfill	& 22\\
Width of ring\tablenotemark{b}~ (kpc) \dotfill		& 0.8\\
R$_{\rm{ring}}$\tablenotemark{b}~ (arcsec)\dotfill	& 47\\
R$_{\rm{ring}}$\tablenotemark{b}~ (kpc)\dotfill		& 1.7\\

\enddata
\tablenotetext{a}{From D array; D = 7.4 Mpc.}
\tablenotetext{b}{From C+D array}

\end{deluxetable}
\clearpage


\clearpage

\clearpage

\clearpage

\clearpage

\clearpage

\clearpage

\clearpage

\clearpage

\clearpage

\clearpage

\clearpage

\clearpage

\clearpage

\clearpage

\clearpage

\end{document}